\documentclass[a4paper,11pt]{article}%
\usepackage{amsmath}
\usepackage{graphicx}
\usepackage{amsbsy}%
\usepackage{amsfonts}%
\usepackage{amssymb}

\begin{document}
\thispagestyle{empty}  \setcounter{page}{0}  \begin{flushright}%

TTP09-04\\SFB/CPP-09-18\\February 2009 \\
\end{flushright}

\vskip        2.2 true cm

\begin{center}
{\huge EDM constraints on flavored CP-violating phases}\\[1.9cm]

\textsc{Lorenzo Mercolli}$^{1}$\textsc{ and Christopher Smith}$^{2}$%
\\[12pt]$^{1}$\textsl{Center for Research and Education in Fundamental Physics,}

\textsl{Institut f\"{u}r Theoretische Physik, Universit\"{a}t Bern, CH-3012
Bern, Switzerland}\\[6pt]

$^{2}$\textsl{Institut f\"{u}r Theoretische Teilchenphysik,}

\textsl{Karlsruhe Institute of Technology, D-76128 Karlsruhe, Germany}\\[1.9cm]

\textbf{Abstract}
\end{center}

\begin{quote}
\noindent\ The CP-violating phenomenology of the MSSM with Minimal Flavor
Violation (MFV) in the lepton sector is revisited. To this end, the most
general parametrizations of the slepton soft-breaking terms are constructed
assuming a seesaw mechanism of type I. After a critical reassessment of how
the CP-symmetry is broken within the MFV framework, all possible CP-violating
phases are introduced. From the strong hierarchy of their contributions to the
Electric Dipole Moments (EDMs), these phases are split into three classes:
flavor-blind, flavor-diagonal and flavor off-diagonal. In particular, the
phases from the neutrino sector belong to the last class; they start to
contribute only at the second order in the mass-insertion approximation and
have thus a negligible effect. It is then shown that to each class of phases
corresponds a unique largely dominant term in the MFV expansion. Numerically,
for a realistic range of MSSM and neutrino parameters, such that
$B(\mu\rightarrow e\gamma)$ does not exceed its experimental bound, the three
types of phases are found to be allowed by the current bound on the electron
EDM, though the next generation of experiments should constrain tightly the
flavor-blind phase. Finally, we relax the MFV hypothesis and show how in the
general MSSM, the MFV operator basis can be used to judge of the naturality of
the slepton soft-breaking terms. \footnotetext[1]{lorenzo@itp.unibe.ch}
\footnotetext[2]{chsmith@particle.uni-karlsruhe.de}\newpage
\end{quote}

\section{Introduction}

In the past few years, the Minimal Flavor Violation (MFV) hypothesis has
emerged as an appealing framework, allowing to reconcile the ever more
constraining low-energy data with the possibility of New Physics below the TeV
scale. Indeed, when the New Physics particles are so light, the absence of
significant deviations with respect to the Standard Model (SM) in $K$ and $B$
observables or in Lepton Flavor Violating (LFV) processes like $\mu\rightarrow
e\gamma$ yields very tight bounds on the absolute value and CP-violating phase
of the flavor-violating couplings. At this stage, one could simply forbid New
Physics to induce any flavor transitions, i.e. force it to be flavor-blind.
However, this seems far too constraining given that flavor transitions do
occur in the SM. By contrast, the MFV hypothesis permits non-trivial flavor
structures for New Physics, but severely limits them by naturally relating
them to those of the SM.

However, there is still some latitude in the precise implementation of MFV,
especially regarding the presence of new CP-violating phases. Indeed, most
approaches try to limit the CP-violating phases in the flavor sector to those
already present in the SM. For instance, in the quark sector, only the CKM
phase is allowed. However, this has several conceptual drawbacks. First, MFV
can be neatly formulated as a symmetry principle, but this procedure leaves
the CP-violating phases completely free. Second, most New Physics models do
contain new unflavored CP-violating phases, so this restriction appears
unnatural and may even be difficult to maintain if the flavor-blind and
flavored sectors can communicate. Finally, as will be discussed in the present
work, the appearance of new CP-violating phases is actually deeply rooted in
the formulation of MFV itself; it could only be naturally avoided by imposing
the CP-symmetry on the whole theory, including on the SM.

Lifting all restrictions on the CP-violating phases, there is a serious danger
of failing to account for the tight bounds on the Electric Dipole Moments
(EDMs), which are flavor-diagonal CP-violating observables. Indeed, for flavor
transitions, one can rely on various hierarchies (like those in the CKM matrix
or in the neutrino masses) to suppress the effects of the new phases, but no
such mechanism is available for flavor-diagonal observables. To be more
specific, the dimension-six effective electromagnetic operator which will
concern us in the following,%
\begin{equation}
H_{eff}=e\frac{\mathcal{C}^{IJ}}{\Lambda^{2}}\bar{\psi}_{R}^{I}\sigma_{\mu\nu
}\psi_{L}^{J}F^{\mu\nu}H_{d}\;, \label{ModIndep}%
\end{equation}
induces the flavor-violating transitions between fermion species $\psi
^{I}\rightarrow\psi^{J}\gamma$ when $\mathcal{C}^{IJ}\neq0$, as well as an EDM
for the fermion $\psi^{I}$ when $\operatorname{Im}\mathcal{C}^{II}\neq0$.
While the couplings $\mathcal{C}^{IJ}$ naturally exhibit a hierarchical
structure for $I\neq J$, inherited from those in the quark and lepton masses
and mixings, the flavor-diagonal $\mathcal{C}^{II}$ are only constrained to be
$\mathcal{O}(1)$ complex numbers by MFV, and can easily induce unacceptably
large EDMs. However, before being drawn to the conclusion that MFV must be
supplemented by some sort of mechanism protecting it from large
flavor-diagonal CP-violating effects, it is worth analyzing the situation in a
precise theoretical setting, to see whether Eq.~(\ref{ModIndep}) gives a fair
account or misses important effects.

This is the purpose of the present paper. We will work within the MSSM, with
neutrino masses generated through a seesaw mechanism of type I, and
investigate whether the constraints from the leptonic EDMs rule out or allow
for the most general CP-violating phases in the slepton sector, as introduced
by MFV. For this program, in the next section, we will start by deriving the
MFV expansions for the slepton mass terms and trilinear couplings and identify
all the CP-violating phases. Then, we will analyze in detail the rationale
behind these additional phases, and classify them according to their impact on
the EDMs. In the third section, we will present the results of our numerical
analyses, showing that all types of MFV phases are allowed, though the planned
improvements in the search for the electron EDM could change that picture. We
will then compare the situation in the MSSM with the model-independent
approach, and show how one can adapt Eq.~(\ref{ModIndep}) so as to reflect the
fact that flavored phases are compatible with experimental bounds on the
leptonic EDMs. In the fourth section, we will move to the general MSSM, and
show how the operator basis constructed in the context of MFV offers a new
perspective on the size of the slepton mass insertions. Finally, our results
are summarized in the conclusion.

\section{The MFV expansion in the slepton sector}

The gauge sector of the MSSM exhibits the $U(3)^{5}$ flavor
symmetry~\cite{ChivukulaG87}, one $U(3)$ factor for each of the quark and
lepton superfields%
\begin{equation}
G_{F}=U(3)^{5}=G_{q}\times G_{\ell}\text{ with }G_{q}\equiv U(3)_{Q}\times
U(3)_{U}\times U(3)_{D}\;,\;G_{\ell}\equiv U(3)_{L}\times U(3)_{E}\;.
\end{equation}
This flavor symmetry is broken by the Yukawa couplings in the superpotential
and by the squark and slepton soft-breaking terms. However, only the former
have been measured precisely through the quark and lepton masses and mixings.
The latter are only constrained to induce limited flavor-mixings when squarks
and sleptons masses are below the TeV scale so as not to violate the many
bounds coming from FCNC and other low-energy observables.

If the mechanism responsible for the breakings of $G_{F}$ by the
superpotential couplings is also behind those due to the soft-breaking terms,
both are necessarily related. The MFV hypothesis is a model-independent
description of such a relationship under the condition of minimality: if the
flavor symmetry is broken just such as to generate the known quark and lepton
masses and mixings, it is then possible to reconstruct the flavor-dependent
soft-breaking couplings~\cite{HallR90,DambrosioGIS02}.

Technically, the minimality statement translates as the existence of only a
limited number of spurion fields breaking $G_{F}$. In the quark sector, the
spurions are taken aligned with the Yukawa couplings%
\begin{equation}
\mathbf{Y}_{u}\sim\left(  \bar{3},3,1\right)  _{G_{q}},\;\;\mathbf{Y}_{d}%
\sim\left(  \bar{3},1,3\right)  _{G_{q}}\;. \label{Yukawas}%
\end{equation}
This immediately makes the MSSM superpotential invariant under $G_{q}$. Then,
the soft-breaking terms are also made formally invariant by writing them
entirely in terms of the Yukawa spurions. For example, the left-squark mass
term takes the form~\cite{HallR90,DambrosioGIS02}%
\begin{equation}
\mathbf{m}_{Q}^{2}=m_{0}^{2}(a_{1}\mathbf{1}+a_{2}\mathbf{Y}_{u}^{\dagger
}\mathbf{Y}_{u}+a_{3}\mathbf{Y}_{d}^{\dagger}\mathbf{Y}_{d}+...)\;.
\end{equation}
To get the corresponding MFV expansions in the slepton sector is the purpose
of the present section.

\subsection{Seesaw spurions}

To account for neutrino masses without introducing unnaturally small neutrino
Yukawa couplings, the MSSM is supplemented with a seesaw mechanism of type
I~\cite{SeeSaw}, i.e. three heavy right-handed neutrino superfields $N$ are
added to its particle content. Though this formally extends the flavor
symmetry group to $G_{F}\times U(3)_{N}$, the $N$ fields never occur at
low-energy and only the spurion combinations which are singlets under
$U(3)_{N}$ are needed~\cite{CiriglianoGIW05}. Expanding in the inverse
right-handed neutrino mass matrix $\mathbf{M}_{R}$, these are
\begin{equation}
\mathbf{Y}_{e}\sim\left(  \bar{3},3\right)  _{G_{\ell}},\;\mathbf{Y}_{\nu
}^{\dagger}\mathbf{Y}_{\nu}\sim\left(  8,1\right)  _{G_{\ell}},\;\mathbf{Y}%
_{\nu}^{T}(\mathbf{M}_{R}^{-1})\mathbf{Y}_{\nu}\sim\left(  \bar{6},1\right)
_{G_{\ell}},\;\mathbf{Y}_{\nu}^{\dagger}(\mathbf{M}_{R}^{-1})^{\ast
}(\mathbf{M}_{R}^{-1})\mathbf{Y}_{\nu}\sim\left(  8,1\right)  _{G_{\ell}%
},\;... \label{Eq5}%
\end{equation}
The $\mathbf{M}_{R}^{-1}$ term corresponds to the Majorana mass term for
left-handed neutrinos. We work in the basis where the charged lepton Yukawa
coupling is diagonal, hence%
\begin{equation}
\mathbf{\Upsilon}_{\nu}\equiv v_{u}\mathbf{Y}_{\nu}^{T}(\mathbf{M}_{R}%
^{-1})\mathbf{Y}_{\nu}=\frac{1}{v_{u}}U^{\ast}\mathbf{m}_{\nu}U^{\dagger
}\;,\;\;\mathbf{Y}_{e}=\frac{\mathbf{m}_{e}}{v_{d}}\;, \label{Bckgrd}%
\end{equation}
with $\mathbf{m}_{e}=\operatorname*{diag}(m_{e},m_{\mu},m_{\tau})$,
$\mathbf{m}_{\nu}=\operatorname*{diag}(m_{\nu1},m_{\nu2},m_{\nu3})$, and
$v_{u,d}=\langle0|H_{u,d}^{0}|0\mathbf{\rangle}$ the two Higgs boson vacuum
expectation values. The mixing matrix $U$ is related to the
Pontecorvo-Maki-Nakagawa-Sakata mixing matrix~\cite{PMNS} as $U=U_{\text{PMNS}%
}.\operatorname*{diag}(1,e^{i\alpha/2},e^{i\beta/2})$, with $U_{\text{PMNS}}$
involving one CP-violating Dirac phase $\gamma$ and where $\alpha$ and $\beta$
are the two Majorana phases.

Provided $\mathbf{M}_{R}$ is sufficiently large, the neutrino Yukawa couplings
can be of $\mathcal{O}\left(  1\right)  $ and so it is the leading spurion
$\mathbf{Y}_{\nu}^{\dagger}\mathbf{Y}_{\nu}$ which will concern us in the
following. As is well known, it is impossible to fix $\mathbf{\mathbf{Y}}%
_{\nu}^{\dagger}\mathbf{\mathbf{Y}}_{\nu}$ unambiguously from neutrino mixing
parameters alone. The arbitrariness can be collected into an unknown complex
orthogonal matrix $\mathbf{R}$~\cite{CasasI01}%
\begin{equation}
\mathbf{Y}_{\nu}^{\dagger}\mathbf{Y}_{\nu}=\frac{1}{v_{u}^{2}}U(\mathbf{m}%
_{\nu}^{1/2})\,\mathbf{R}^{\dagger}\mathbf{\,M}_{R}\mathbf{\,R\,}%
(\mathbf{m}_{\nu}^{1/2})U^{\dagger}\;.
\end{equation}
As discussed in Ref.~\cite{CiriglianoIP06}, the right-handed neutrinos are, to
a good approximation, degenerate when MFV is enforced at the seesaw scale.
Taking $\mathbf{M}_{R}=M_{R}\mathbf{1}$, the above spurion simplifies
to~\cite{PascoliPY03}%
\begin{equation}
\mathbf{\mathbf{Y}}_{\nu}^{\dagger}\mathbf{\mathbf{Y}}_{\nu}=\frac{M_{R}%
}{v_{u}^{2}}U(\mathbf{m}_{\nu}^{1/2})\,e^{2i\mathbf{\Phi}}\mathbf{\,}%
(\mathbf{m}_{\nu}^{1/2})U^{\dagger}\;, \label{Spurion}%
\end{equation}
with the matrix $\mathbf{\Phi}^{ij}=\varepsilon^{ijk}\phi_{k}$ involving three
(real) parameters $\phi_{1,2,3}$, on which little is known. They affect the
size of the CP-conserving entries in $\mathbf{\mathbf{Y}}_{\nu}^{\dagger
}\mathbf{\mathbf{Y}}_{\nu}$ and induce CP-violating imaginary parts, since%
\begin{equation}
e^{2i\mathbf{\Phi}}=e^{i\mathbf{\Phi}}.e^{i\mathbf{\Phi}}=\mathbf{1}%
-\frac{2\sinh^{2}r}{r^{2}}\mathbf{\Phi.\Phi}+i\frac{\sinh2r}{r}\mathbf{\Phi
,\;\;}r\equiv\sqrt{\phi_{1}^{2}+\phi_{2}^{2}+\phi_{3}^{2}}\;. \label{phii}%
\end{equation}
Note that if $\phi_{i}=0$, $\mathbf{\mathbf{Y}_{\nu}^{\dagger}\mathbf{Y}_{\nu
}}$ becomes independent of the Majorana phases $\alpha$ and $\beta$. If the
Dirac phase $\gamma$ also vanishes, one simply has $\mathbf{\mathbf{Y}}_{\nu
}^{\dagger}\mathbf{\mathbf{Y}}_{\nu}=(M_{R}/v_{u})\mathbf{\Upsilon}_{\nu}$.

\subsection{Reparametrizations and algebraic reductions}

The MSSM is made formally invariant under $G_{\ell}$ by writing the
soft-breaking terms $\mathbf{m}_{L}^{2}$, $\mathbf{m}_{E}^{2}$, and
$\mathbf{A}_{e}$ in terms of the two spurions $\mathbf{Y}_{e}$ and
$\mathbf{\mathbf{Y}}_{\nu}^{\dagger}\mathbf{\mathbf{Y}}_{\nu}$. Let us first
consider a generic operator $\mathbf{Q}$ transforming as an octet under
$U(3)_{L}$. In full generality, it can be parametrized as an infinite series
of products of powers of
\begin{equation}
\mathbf{A}\equiv\mathbf{Y}_{e}^{\dagger}\mathbf{Y}_{e},\;\mathbf{B}%
\equiv\mathbf{Y}_{\nu}^{\dagger}\mathbf{Y}_{\nu}\;,
\end{equation}
also transforming as octets under $U(3)_{L}$, as%
\begin{equation}
\mathbf{Q}=\sum_{i,j,k,...=0,1,2,...}z_{ijk...}\mathbf{A}^{i}\mathbf{B}%
^{j}\mathbf{A}^{k}...\;, \label{Generalz}%
\end{equation}
for some appropriate coefficients $z_{ijk...}$, a priori all complex. This
series can be partially resummed using the Cayley-Hamilton identity for a
generic $3\times3$ matrix $\mathbf{X}$%
\begin{equation}
\mathbf{X}^{3}-\langle\mathbf{X\rangle X}^{2}+\tfrac{1}{2}\mathbf{X}%
(\langle\mathbf{X\rangle}^{2}-\langle\mathbf{X}^{2}\mathbf{\rangle)}%
-\det\mathbf{X}=0\;,\; \label{CH1}%
\end{equation}
where $\langle\mathbf{X\rangle}$ denotes the trace of $\mathbf{X}$, as well as
identities derived from it (see e.g. Ref.~\cite{ColangeloNS08}), leaving the
operator $\mathbf{Q}$ with a finite number of terms:%
\begin{align}
\mathbf{Q}  &  =y_{1}\mathbf{1}+y_{2}\mathbf{A}+y_{3}\mathbf{B}+y_{4}%
\mathbf{A}^{2}+y_{5}\mathbf{B}^{2}+y_{6}\mathbf{AB}+y_{7}\mathbf{BA}%
+y_{8}\mathbf{ABA}+y_{9}\mathbf{BA}^{2}+y_{10}\mathbf{BAB}.\nonumber\\
&  \;\;\;\;+y_{11}\mathbf{AB}^{2}+y_{12}\mathbf{ABA}^{2}+y_{13}\mathbf{A}%
^{2}\mathbf{B}^{2}+y_{14}\mathbf{B}^{2}\mathbf{A}^{2}+y_{15}\mathbf{B}%
^{2}\mathbf{AB}+y_{16}\mathbf{AB}^{2}\mathbf{A}^{2}+y_{17}\mathbf{B}%
^{2}\mathbf{A}^{2}\mathbf{B}\;. \label{MFVoctet}%
\end{align}
The only non-trivial reduction is that for the term $\mathbf{B}^{2}%
\mathbf{ABA}^{2}$, related to the leptonic Jarlskog
invariant~\cite{Jarlskog85}, and can be done starting with $\mathbf{X}%
=[\mathbf{A},\mathbf{B}]$ in Eq.~(\ref{CH1}).

Still using only the Cayley-Hamilton identities, and thanks to the hermiticity
of $\mathbf{Y}_{e}^{\dagger}\mathbf{Y}_{e}$ and $\mathbf{Y}_{\nu}^{\dagger
}\mathbf{Y}_{\nu}$, we can manipulate the expansion~(\ref{MFVoctet}) to write
$\mathbf{Q}$ entirely in terms of hermitian operators as%
\begin{align}
\mathbf{Q}  &  =x_{1}\mathbf{1}+x_{2}\mathbf{A}+x_{3}\mathbf{B}+x_{4}%
\mathbf{A}^{2}+x_{5}\mathbf{B}^{2}+x_{6}i[\mathbf{A},\mathbf{B}]+x_{7}%
\{\mathbf{A},\mathbf{B}\}+x_{8}\mathbf{ABA}\nonumber\\
&  \;\;\;\;+x_{9}i[\mathbf{B},\mathbf{A}^{2}]+x_{10}\mathbf{BAB}%
+x_{11}i[\mathbf{A},\mathbf{B}^{2}]+x_{12}\mathbf{BA}^{2}\mathbf{B}%
+x_{13}i[\mathbf{A}^{2},\mathbf{B}^{2}]\nonumber\\
&  \;\;\;\;+x_{14}i(\mathbf{ABA}^{2}-\mathbf{A}^{2}\mathbf{BA})+x_{15}%
i(\mathbf{B}^{2}\mathbf{AB}-\mathbf{BAB}^{2})\nonumber\\
&  \;\;\;\;+x_{16}i(\mathbf{AB}^{2}\mathbf{A}^{2}-\mathbf{A}^{2}\mathbf{B}%
^{2}\mathbf{A})+x_{17}i(\mathbf{B}^{2}\mathbf{A}^{2}\mathbf{B}-\mathbf{BA}%
^{2}\mathbf{B}^{2})\;. \label{MFVoctetH}%
\end{align}
For the last four operators, the corresponding ''+'' hermitian combinations
are fully reducible. For example, the combination $\mathbf{ABA}^{2}%
+\mathbf{A}^{2}\mathbf{BA}$ is entirely absorbed into lower-order terms using%
\begin{align}
\mathbf{A}^{2}\mathbf{BA}+\mathbf{ABA}^{2}  &  =\mathbf{ABA}\left\langle
\mathbf{A}\right\rangle +\mathbf{A}^{2}\left\langle \mathbf{BA}\right\rangle
-\tfrac{1}{6}\mathbf{B(}\langle\mathbf{\mathbf{A}}\rangle^{3}+2\langle
\mathbf{A}^{3}\rangle-3\langle\mathbf{A}\rangle\langle\mathbf{A}^{2}%
\rangle)+\mathbf{A}\left\langle \mathbf{BA}^{2}\right\rangle \nonumber\\
&  \;\;\;\;-\mathbf{A}\left\langle \mathbf{BA}\right\rangle \left\langle
\mathbf{A}\right\rangle +\tfrac{1}{2}\left\langle \mathbf{BA}\right\rangle
(\left\langle \mathbf{A}\right\rangle ^{2}-\left\langle \mathbf{A}%
^{2}\right\rangle )+\left\langle \mathbf{A}^{2}\mathbf{BA}\right\rangle
-\left\langle \mathbf{A}\right\rangle \left\langle \mathbf{BA}^{2}%
\right\rangle \;,
\end{align}
which is derived from Eq.~(\ref{CH1}).

The most general expansions for $\mathbf{m}_{L}^{2}$, $\mathbf{m}_{E}^{2}$ and
$\mathbf{A}_{e}$ are directly obtained from the expansion of $\mathbf{Q}$ as
\begin{equation}
\mathbf{m}_{L}^{2}=m_{0}^{2}\mathbf{Q},\;\;\mathbf{m}_{E}^{2}=m_{0}%
^{2}(\mathbf{1}+\mathbf{Y}_{e}\mathbf{QY}_{e}^{\dagger}),\;\;\mathbf{A}%
_{e}=A_{0}\mathbf{Y}_{e}\mathbf{Q}\;, \label{MGE}%
\end{equation}
where it is understood that the MFV coefficients for $\mathbf{m}_{L}^{2}$ and
$\mathbf{m}_{E}^{2}$ are real, since these mass terms are hermitian, while
those for $\mathbf{A}_{e}$ can be complex. The mass parameters $m_{0}$ and
$A_{0}$ set the supersymmetry breaking scale. These expansions are the most
general parametrizations of the slepton soft-breaking terms in presence of a
type I seesaw mechanism. In particular, if the neutrino sector communicates to
the slepton sector through RGE effects~\cite{BorzumatiM86}, they correspond to
the most general form one could attain starting with universal slepton
soft-breaking terms, up to arbitrary high orders. Finally, if one allows for
$U(1)_{L}$ and $U(1)_{E}$-breaking terms, operators involving contractions of
$\mathbf{Q}$ with the $SU(3)_{L,E}$ Levi-Civita tensors can be constructed for
$\mathbf{A}_{e}$. These are a priori smaller and will not be included
here~\cite{NikolidakisS07}.

It should be also remarked that all these developments can be immediately
applied to the quark sector by substituting $\mathbf{A}\equiv\mathbf{Y}%
_{e}^{\dagger}\mathbf{Y}_{e}\rightarrow\mathbf{Y}_{d}^{\dagger}\mathbf{Y}_{d}$
and $\mathbf{B}\equiv\mathbf{Y}_{\nu}^{\dagger}\mathbf{Y}_{\nu}\rightarrow
\mathbf{Y}_{u}^{\dagger}\mathbf{Y}_{u}$ in Eqs.~(\ref{Generalz}),
(\ref{MFVoctet}) or (\ref{MFVoctetH}). The corresponding expansions for
$\mathbf{m}_{Q}^{2}$, $\mathbf{m}_{U}^{2}$, $\mathbf{m}_{D}^{2}$,
$\mathbf{A}_{u}$ and $\mathbf{A}_{d}$ have been derived in
Ref.~\cite{ColangeloNS08}.

\subsection{MFV expansions and numerical reductions}

The expansions (\ref{MGE}) are fully general and do not correspond to MFV yet.
Indeed, any matrix can be expanded in the basis (\ref{MFVoctetH}), which is
thus more a reparametrization than an expansion. However, projecting an
arbitrary matrix, the coefficients $x_{i}$ in general span several orders of
magnitudes because the spurion operators in $\mathbf{Q}$ are nearly
aligned~\cite{ColangeloNS08,NikoPhD}.

In a more realistic framework, one would expect the coefficients $z_{ijk...}$
of Eq.~(\ref{Generalz}) to be at most $\mathcal{O}(1)$ complex numbers. Then,
the coefficients $y_{i}$ of Eq.~(\ref{MFVoctet}) and $x_{i}$ of
Eq.~(\ref{MFVoctetH}) are also of $\mathcal{O}(1)$, since the Cayley-Hamilton
identities never generate large numerical coefficients (all the traces of
$\mathbf{A}$, $\mathbf{B}$ and combinations are $\mathcal{O}(1)$ or smaller).
This constraint of naturalness on the size of the expansion coefficients is
the essence of the MFV hypothesis. The MFV expansions for the slepton
soft-breaking terms differ from a mere reparametrization only in the initial
constraints imposed on the coefficients.

Still, once the expansion coefficients are assumed to be at most of
$\mathcal{O}(1)$, an additional reduction of the number of terms is possible.
It stems from the large mass hierarchy between the charged leptons and,
contrary to the Cayley-Hamilton reduction, is only approximate. Following
Ref.~\cite{ColangeloNS08}, this hierarchy is accounted for through the
identification $(\mathbf{Y}_{e}^{\dagger}\mathbf{Y}_{e})^{2}\approx y_{\tau
}^{2}\mathbf{Y}_{e}^{\dagger}\mathbf{Y}_{e}$. Of course, no similar identity
exists for the neutrino spurion. Getting rid of all the terms involving
$\mathbf{A}^{2}$, the MFV series reduces to%
\begin{align}
\mathbf{Q}  &  =c_{1}\mathbf{1}+c_{2}\mathbf{A}+c_{3}\mathbf{B}+c_{4}%
\mathbf{B}^{2}+c_{5}\{\mathbf{A},\mathbf{B}\}+c_{6}\mathbf{BAB}\nonumber\\
&  \;\;\;\;+c_{7}i[\mathbf{A},\mathbf{B}]+c_{8}i[\mathbf{A},\mathbf{B}%
^{2}]+c_{9}i(\mathbf{B}^{2}\mathbf{AB}-\mathbf{BAB}^{2})\;, \label{HermQ}%
\end{align}
where the nine coefficients $c_{i}$ are $\mathcal{O}(1)$ complex numbers, or
if $\mathbf{Q}$ is hermitian, nine $\mathcal{O}(1)$ real numbers. Plugging
these expressions into Eq.~(\ref{MGE}), the final MFV expansions for the
slepton soft-breaking terms are (remember $\mathbf{A}\equiv\mathbf{Y}%
_{e}^{\dagger}\mathbf{Y}_{e}$ and $\mathbf{B}\equiv\mathbf{Y}_{\nu}^{\dagger
}\mathbf{Y}_{\nu}$):
\begin{subequations}
\label{MFV}%
\begin{align}
\mathbf{m}_{L}^{2}  &  =m_{0}^{2}(a_{1}\mathbf{1}+a_{2}\mathbf{A}%
+a_{3}\mathbf{B}+a_{4}\mathbf{B}^{2}+a_{5}\{\mathbf{A},\mathbf{B}%
\}+a_{6}\mathbf{BAB}\nonumber\\
&  \;\;\;\;\;\;\;\;\;\;\;+ib_{1}[\mathbf{A},\mathbf{B}]+ib_{2}[\mathbf{A}%
,\mathbf{B}^{2}]+ib_{3}(\mathbf{B}^{2}\mathbf{AB}-\mathbf{BAB}^{2}))\;,\\
\mathbf{m}_{E}^{2}  &  =m_{0}^{2}(a_{7}\mathbf{1}+\mathbf{Y}_{e}%
(a_{8}\mathbf{1}+a_{9}\mathbf{B}+a_{10}\mathbf{B}^{2}+a_{11}\{\mathbf{A}%
,\mathbf{B}\}+a_{12}\mathbf{BAB}\nonumber\\
&  \;\;\;\;\;\;\;\;\;\;\;+ib_{4}[\mathbf{A},\mathbf{B}]+ib_{5}[\mathbf{A}%
,\mathbf{B}^{2}]+ib_{6}(\mathbf{B}^{2}\mathbf{AB}-\mathbf{BAB}^{2}%
))\mathbf{Y}_{e}^{\dagger})\;,\\
\mathbf{A}_{e}  &  =A_{0}\mathbf{Y}_{e}(c_{1}\mathbf{1}+c_{2}\mathbf{A}%
+c_{3}\mathbf{B}+c_{4}\mathbf{B}^{2}+c_{5}\{\mathbf{A},\mathbf{B}%
\}+c_{6}\mathbf{BAB}\nonumber\\
&  \;\;\;\;\;\;\;\;\;\;\;\;\;\;+d_{1}i[\mathbf{A},\mathbf{B}]+d_{2}%
i[\mathbf{A},\mathbf{B}^{2}]+d_{3}i(\mathbf{B}^{2}\mathbf{AB}-\mathbf{BAB}%
^{2}))\;.
\end{align}
Altogether, the 18 real $a_{i}$ and $b_{i}$, and the nine complex $c_{i}$ and
$d_{i}$ MFV coefficients sum up to 36 free real parameters, in addition to the
dimensionful SUSY-breaking scale parameters $m_{0}$ and $A_{0}$. These
expansions span the whole space of the complex (hermitian) matrices for
complex (real) coefficients, but let us stress once more that expanding an
arbitrary matrix in those bases generates huge coefficients, in contradiction
with the MFV hypothesis~\cite{ColangeloNS08,NikoPhD}.

\subsection{CP-violation under the MFV hypothesis}

There are two possible sources of CP-violation in the expansions~(\ref{MFV}).
First, the spurion $\mathbf{\mathbf{Y}}_{\nu}^{\dagger}\mathbf{\mathbf{Y}%
}_{\nu}$ involves several CP-violating parameters: the Dirac phase $\gamma$,
the two Majorana phases $\alpha$ and $\beta$, and the three parameters
$\phi_{i}$. The second source are the MFV coefficients themselves. The $b_{i}$
are all purely CP-violating, while the $c_{i}$ and $d_{i}$ can have a
CP-violating component. The goal of the present section is to characterize
these flavored phases --i.e., coming from the slepton masses and trilinear
terms of Eq.~(\ref{MFV})--, by organizing them into three classes:
flavor-blind, flavor diagonal and flavor off-diagonal. Before this, we first
discuss in detail why CP-violating phases have to be allowed for the MFV coefficients.

\paragraph{Necessity for CP-violating coefficients.}

The MFV coefficients are not required by the $U(3)^{5}$ symmetry to be
CP-conserving, but it is actually a matter of consistency to allow them to
violate CP. Looking back at the reduction from the general expansion
(\ref{Generalz}) down to (\ref{MFVoctet}) or (\ref{MFVoctetH}), traces of
combinations of the spurions are absorbed into the coefficients through the
use of Cayley-Hamilton identities. These traces can be complex when the
spurion $\mathbf{\mathbf{Y}}_{\nu}^{\dagger}\mathbf{\mathbf{Y}}_{\nu}$
involves CP-violating phases, so the coefficients have to be allowed to be
complex. In fact, independently of Cayley-Hamilton identities, MFV
coefficients are always understood to include (potentially complex) traces of
products of the spurions, up to any order, so even the $z_{ijk...}$ of
Eq.~(\ref{Generalz}) should not be taken real. The same requirement arises if
the MFV expansion is set at a different scale: the necessity for complex
coefficients appears when the radiative corrections to the soft-breaking terms
are projected back on the standard expansions using the Cayley-Hamilton
identities\cite{ColangeloNS08}. In this case, one also sees that it would be
inconsistent to partially account for CP-violating phases, for example by
allowing complex MFV coefficients for $\mathbf{A}_{e}$ but not $\mathbf{m}%
_{L,E}^{2}$, because they are linked through the RGE\cite{MartinV94}.

Phenomenologically, it would nevertheless be useful to have at hand a
well-defined CP-limit for the MFV coefficients, but this is actually
ill-defined: it depends on the operator basis chosen. Naively, one may think
that taking all the MFV coefficients real corresponds to the CP-limit, but
this is obviously not correct since, looking at Eq.~(\ref{MFVoctet}) and
(\ref{MFVoctetH}),%
\end{subequations}
\begin{equation}
\mathbf{Q}_{\operatorname{Im}y_{i}\rightarrow0}\nLeftrightarrow\mathbf{Q}%
_{\operatorname{Im}x_{i}\rightarrow0}\;.
\end{equation}
So long as the spurion is complex, there is no reason to prefer one limit over
the other. On the other hand, if the CP-limit is enforced also for the spurion
$\mathbf{\mathbf{Y}}_{\nu}^{\dagger}\mathbf{\mathbf{Y}}_{\nu}$, the CP-limit
for $\mathbf{Q}$ is obtained setting $\operatorname{Im}y_{i}\rightarrow0$ or
$\operatorname{Im}x_{1-8}\rightarrow0$, $\operatorname{Re}x_{9-17}%
\rightarrow0$, since then $\mathbf{Q}$ becomes real. These two limits are
further equivalent when $\mathbf{\mathbf{Y}}_{\nu}^{\dagger}\mathbf{\mathbf{Y}%
}_{\nu}$ is complex, because Eq.~(\ref{MFVoctet}) and (\ref{MFVoctetH})
essentially differ by some $i$ factors and some real rearrangements, but this
needs not be the case. For example, the CP-limit is obtained from
Eq.~(\ref{Generalz}) and (\ref{MFVoctet}) by setting $\operatorname{Im}%
z_{i}\rightarrow0$ and $\operatorname{Im}y_{i}\rightarrow0$ when
$\mathbf{\mathbf{Y}}_{\nu}^{\dagger}\mathbf{\mathbf{Y}}_{\nu}$ is real.
However, when $\mathbf{\mathbf{Y}}_{\nu}^{\dagger}\mathbf{\mathbf{Y}}_{\nu}$
is complex, these same limits are no longer equivalent,
\begin{equation}
\mathbf{Q}_{\operatorname{Im}z_{i}\rightarrow0}\nLeftrightarrow\mathbf{Q}%
_{\operatorname{Im}y_{i}\rightarrow0}\;,
\end{equation}
because in going from (\ref{Generalz}) to (\ref{MFVoctet}), some $y_{i}$ have
absorbed complex traces. Therefore, we arrive at the conclusion that the
CP-limit for the MFV coefficient is basis-dependent, and thus again that as a
matter of principle, it makes no sense to impose a CP-limit on the MFV
coefficients while allowing the spurions to be CP-violating.

In the present case, the spurions $\mathbf{\mathbf{Y}}_{\nu}^{\dagger
}\mathbf{\mathbf{Y}}_{\nu}$ and $\mathbf{\mathbf{Y}}_{e}^{\dagger
}\mathbf{\mathbf{Y}}_{e}$ are both hermitian, hence all the complex traces can
be reduced to the Jarlskog invariant using the Cayley-Hamilton identities%
\begin{equation}
J\equiv\operatorname{Im}\langle(\mathbf{\mathbf{Y}}_{\nu}^{\dagger
}\mathbf{\mathbf{Y}}_{\nu}\mathbf{\mathbf{)}}^{2}\mathbf{\mathbf{Y}}%
_{e}^{\dagger}\mathbf{\mathbf{Y}}_{e}\mathbf{\mathbf{\mathbf{Y}}}_{\nu
}^{\dagger}\mathbf{\mathbf{\mathbf{Y}}}_{\nu}(\mathbf{\mathbf{\mathbf{Y}}}%
_{e}^{\dagger}\mathbf{\mathbf{\mathbf{Y}}}_{e})^{2}\rangle=\frac{i}{2}%
\det[\mathbf{\mathbf{Y}}_{\nu}^{\dagger}\mathbf{\mathbf{Y}}_{\nu
},\mathbf{\mathbf{Y}}_{e}^{\dagger}\mathbf{\mathbf{Y}}_{e}]\neq0\;.
\label{Jarlskog}%
\end{equation}
This invariant is very small, and thus the ambiguities in the imaginary parts
of the MFV coefficients discussed before are also very small. The potentially
large CP-violating effects from the $\phi_{i}$'s cannot enhance $J$ much
because these parameters affect both the real and imaginary parts of the
entries of $\mathbf{\mathbf{Y}}_{\nu}^{\dagger}\mathbf{\mathbf{Y}}_{\nu}$, see
Eq.~(\ref{phii}), and the perturbativity bound $|\mathbf{\mathbf{Y}}_{\nu
}^{\dagger}\mathbf{\mathbf{Y}}_{\nu}|\lesssim1$ fails much before $J$ could
reach even the percent level. However, the crucial point is that even if $J$
is very small, its non-zero value implies that the CP-symmetry cannot be
defined separately for the MFV coefficients and for the spurions. It can
therefore not be used to enforce naturally some reality conditions on the MFV
coefficients. Without any acting symmetry, enforcing such a condition brings
instead a fine-tuning problem. This goes against the very principle of MFV
which is to restore naturality in the flavor sector.

Of course, this fine-tuning of the CP-violating phases could have a dynamical
explanation, in which case we proved above that it is numerically stable under
a change of MFV operator basis and under the RG evolution. For instance, in
the quark sector, it was shown in Ref.~\cite{ColangeloNS08} that imposing MFV
at the GUT scale and running it down, the MFV coefficients exhibit a quasi
fixed-point behavior, with their imaginary parts running towards negligible
values. However, this peculiar behavior can be traced back to the fast
evolution of the flavor-blind parameters --essentially evolving like the
gluino mass--, and MFV coefficients in the slepton sector are not expected to
exhibit such a behavior. Therefore, lacking a dynamical mechanism able to
enforce a fine-tuning of the slepton MFV coefficients, we will allow them to
be $\mathcal{O}(1)$ complex numbers. Ultimately, it is the comparison with
experiment which will tell us if this setting is viable or not.

Finally, it should be mentioned that the Jarlskog invariant (\ref{Jarlskog})
is not always the only possible one. For instance, when the right-handed
neutrinos are not degenerate, their successive decouplings at different scales
give rise to an additional spurion, from which a larger Jarlskog invariant can
be constructed~\cite{NonDegenerateMR}. In such a case, fine-tuning the MFV
coefficients to CP-conserving values is not longer numerically stable, and any
MFV implementation must then necessarily involve complex coefficients. These
effects will be left for subsequent studies, and we will concentrate here on
the minimal spurion content only.

\paragraph{Flavor-blind phases and phase-convention dependences.}

For the slepton mass terms $\mathbf{m}_{L}^{2}$ and $\mathbf{m}_{E}^{2}$, MFV
only introduces relative phases between coefficients because of the
hermiticity constraint; specifically, all the $a_{i}$ and $b_{i}$ are taken
real. For the trilinear term $\mathbf{A}_{e}$, all the $c_{i}$ and $d_{i}$
coefficients can be complex. Their phases thus have to be defined relative to
those in the gauge sector of the MSSM. Since the latter are flavor-blind
(i.e., not related to $U(3)^{5}$ breakings), it appears that MFV introduces a
relative flavor-blind phase in $\mathbf{A}_{e}$. In other words, parts of the
$\mathbf{A}_{e}$ phases can be moved into the flavor-blind parameters of the
MSSM (gaugino masses, $\mu$ term,...) by changing the phase conventions (see
e.g. Ref.~\cite{AbelKL01}).\footnote{Strictly speaking, the flavor-blind phase
is nevertheless ``flavored'' since $\mathbf{A}_{e}$ has no purely flavor-blind
component. Indeed, without the spurion $\mathbf{Y}_{e}$, $\mathbf{A}_{e}$ is
forbidden in MFV.}

At this stage, there are two possible points of view. One can simply accept
that all the phases in $\mathbf{A}_{e}$ are free and well-defined once those
in the rest of the MSSM are fixed. The $c_{i}$ and $d_{i}$ are then allowed to
take any $\mathcal{O}(1)$ complex values, provided they are compatible with
EDM constraints. The second point of view is to consider only relative phases
between MFV coefficients, and to assume that the flavor-blind relative phase
between the MFV expansion and the rest of the MSSM is fixed by some other,
unknown mechanism. Indeed, it is well known that the flavor-blind phases of
the gaugino masses or the $\mu$ term can easily lead to much too large EDMs if
sparticles are not very heavy (the so-called MSSM CP problem~\cite{CPproblem}%
). MFV does not constrain these phases, it is up to some other mechanism to
restrict their sizes.\footnote{Alternatively, cancellations between the MSSM
contributions to the EDM could be at play, see e.g. Refs.\cite{Cancellation}.}
This unknown mechanism could then also constrain the overall flavor-blind
phase of $\mathbf{A}_{e}$. In the present work, we will not take this second
point of view, because we want to quantify the size of the effects induced by
natural phases in $\mathbf{A}_{e}$ (in the MFV sense), but this possibility
should be kept in mind. Further, the phase of the other relevant parameters
for the EDMs, $\mu$, $M_{1}$ and $M_{2}$, will be set to zero since they are
beyond the reach of MFV.

\paragraph{Three types of CP-violating phases for leptonic EDMs.}

Given the above provisions, we distinguish three types of CP-violating phases
according to their effects on leptonic EDMs.

The first type is the \textit{flavor-blind} one. Looking back at the general
expansion (\ref{Generalz}), it seems natural to identify it as the phase of
$z_{0}$, leaving to the physics behind the MFV expansion the task to generate
the relative phases of all the other coefficients $z_{ijk...}$, whose
operators explicitly break the flavor symmetry. This identification does not
immediately permit to pinpoint the flavor-blind phase in the
expansion~(\ref{MFV}), because of the systematic use of Cayley-Hamilton
identities; the initial constraint on $z_{0}$ is only passed to $c_{1}$ when
the spurions are sufficiently suppressed. This is however the case in a large
portion of the parameter space, as we will see in the next section. We
therefore call $\arg c_{1}$ (or, with a small abuse of language,
$\operatorname{Im}c_{1}$) the flavor-blind phase. When present, the
contribution of this phase to the EDMs dominates.

All the other CP-violating phases can be split into two classes according to
the order in the Mass-Insertion Approximation (MIA) at which they start to
contribute to the leptonic EDMs. This is the place where having written the
MFV series entirely in terms of hermitian operators becomes important. Indeed,
the diagonal entries of these operators are automatically real and their
contributions to the EDMs are relegated to the second order in the MIA. In
other words, the EDMs are effectively shielded from direct effects from the
CP-violating phases occurring in $\mathbf{\mathbf{Y}}_{\nu}^{\dagger
}\mathbf{\mathbf{Y}}_{\nu}$.\footnote{This holds also for $\mathbf{A}_{e}$,
which is the product of hermitian operators with $\mathbf{\mathbf{Y}}_{e}$,
real and diagonal in the basis (\ref{Bckgrd}).} This situation is similar to
the flavor off-diagonal CP-violation scenarios discussed e.g. in
Ref.~\cite{AbelKL01}, even though $\mathbf{A}_{e}$ itself is not hermitian here.

The parameters $\operatorname{Im}c_{i}$ and $\operatorname{Im}d_{3}$ induce
leading-order MIA effects, and are thus called \textit{flavor-diagonal
phases}. On the other hand, $b_{i}$, $d_{1,2}$ and all the $\mathbf{\mathbf{Y}%
}_{\nu}^{\dagger}\mathbf{\mathbf{Y}}_{\nu}$ spurion phases contribute only at
the second order in the MIA, and are called\textit{ flavor off-diagonal
phases}. Note that $d_{1,2}$ do not contribute to the leading order in the MIA
because the diagonal entries of $[\mathbf{A},\mathbf{B}]$ and $[\mathbf{A}%
,\mathbf{B}^{2}]$ always vanish (this can be traced back to the fact that
those of $\mathbf{AB}$ and $\mathbf{AB}^{2}$ are purely real). On the other
hand, $d_{3}$ does have non-vanishing diagonal entries but, being of third
order in $\mathbf{\mathbf{Y}}_{\nu}^{\dagger}\mathbf{\mathbf{Y}}_{\nu}$, they
are too suppressed to play any role.

\section{MFV phases and leptonic observables}

All the CP-violating phases occurring in the MFV expansions induce corrections
to the leptonic EDMs. The goal of the present section is to estimate the order
of magnitude of these effects for each type of phases, i.e. flavor-blind,
flavor diagonal and flavor off-diagonal, and to see whether they are
compatible with the experimental bounds. Also, we want to characterize the
contribution of each type of phases by finding which of the MFV operators are
relevant, and which are always negligible.

Instead of a full scan over the MSSM parameter space, our strategy will be to
first identify a reasonable range of parameters, i.e. such that LFV
transitions (essentially $\mu\rightarrow e\gamma$) satisfy their experimental
bounds, and then to compute the EDMs over this range. In this way, the loose
correlation between these two types of leptonic observables is fully
exploited, and the effect of their different dependences on MFV coefficients
and neutrino parameters can be probed most thoroughly.

First, in the next subsection, we give the relevant expressions for the
one-loop contributions to the EDM and LFV transitions in the MSSM, together
with the current experimental bounds on these observables. Also, the various
experimental inputs, as well as the ranges of variation we allow for the
unknown seesaw and MSSM parameters, are described.

\subsection{Preliminaries: Electromagnetic operator and parameter input values}

The supersymmetric contributions can be collected into a single dimension-five
effective operator%
\begin{equation}
H_{eff}=e\mathcal{M}^{IJ}\bar{\psi}_{L}^{I}\sigma_{\mu\nu}\psi_{R}^{J}%
F^{\mu\nu}+h.c.
\end{equation}
The flavor diagonal parts $\mathcal{M}^{II}$ are directly related to the EDMs
and Magnetic Dipole Moments (MDMs), $d_{I}/e=2\operatorname{Im}\mathcal{M}%
^{II}$ and $a_{I}=4m_{e^{I}}\operatorname{Re}\mathcal{M}^{II}$, respectively,
while the $\ell^{I}\rightarrow\ell^{J}\gamma$ transitions arise from the
off-diagonal parts $\mathcal{M}^{IJ}$.

For our purposes, it is sufficient to consider only the neutralino and
chargino one-loop contributions. We thus neglect Barr-Zee type
contributions~\cite{BarrZee} as well as contributions arising from the
$\tan\beta$-enhanced non-holomorphic corrections to the Yukawa
couplings~\cite{Paradisi}. We take the results of Ref.~\cite{MasinaS02}, with
LFV transitions to first order in the MIA,%
\begin{align}
B\left(  \ell^{I}\rightarrow\ell^{J}\gamma\right)   &  =\frac{3\alpha}{2\pi
}\tan^{4}\theta_{W}\times B\left(  \ell^{I}\rightarrow\ell^{J}\bar{\nu}^{J}%
\nu^{I}\right)  \times\frac{M_{W}^{4}\,M_{1}^{2}\tan^{2}\beta}{|\mu|^{2}%
}\nonumber\\
&  \;\;\;\;\times\left(  \left|  \delta_{LL}^{JI}\mathcal{F}_{1}+\delta
_{LR}^{JI}\frac{m_{R}m_{L}}{\mu m_{e}^{I}\tan\beta}\mathcal{F}_{2}\right|
^{2}+\left|  \delta_{RR}^{JI}\mathcal{F}_{3}+\delta_{RL}^{JI}\frac{m_{R}m_{L}%
}{\mu^{\ast}m_{e}^{I}\tan\beta}\mathcal{F}_{2}\right|  ^{2}\right)  \;,
\label{LFV}%
\end{align}
and the lepton EDMs and MDMs up to second order,%
\begin{align}
\frac{a_{I}}{m_{e}^{I}}+2i\frac{d_{I}}{e}  &  =\frac{\alpha M_{1}}{4\pi
|\mu|^{2}\cos^{2}\theta_{W}}\left[  m_{e}^{I}\left(  \mu\tan\beta
\mathcal{F}_{4}-A_{e}^{I\ast}\mathcal{F}_{5}\right)  +m_{R}m_{L}\left(
\delta_{LR}^{IK}\delta_{RR}^{KI}\mathcal{F}_{7}+\delta_{LL}^{IK}\delta
_{LR}^{KI}\mathcal{F}_{8}\right)  \frac{{}}{{}}\right. \nonumber\\
&  \;\;\;\;\;\;\;\;\;\;\;\;\;\;\;\;\;\;\;\left.  +m_{e^{K}}\left(  \delta
_{LL}^{IK}\delta_{RR}^{KI}\left(  \mu\tan\beta-A_{e}^{K\ast}\right)
+\delta_{LR}^{IK}\delta_{LR}^{KI}\left(  \mu^{\ast}\tan\beta-A_{e}^{K}\right)
\right)  \mathcal{F}_{6}+...\frac{{}}{{}}\right]  \;. \label{EDM}%
\end{align}
In these expressions, $m_{e}^{I}$ is the $\ell^{I}$ mass, $m_{L}$ and $m_{R}$
are the average left and right slepton masses used in defining the
mass-insertions $\delta_{LL}$, $\delta_{RR}$ and $\delta_{LR}$, while
$A_{e}^{I}$ are defined as the diagonal entries of $(\mathbf{Y}_{e}%
)^{-1}\mathbf{A}_{e}$. The loop functions $\mathcal{F}_{i}$, all of
mass-dimension $M^{-2}$, can be found in Ref.~\cite{MasinaS02}. In the actual
computation, we use also the full one-loop results of Ref.~\cite{Loop}. We
checked, as stated in Ref.~\cite{MasinaS02}, that the MIA is excellent over
most of the parameter space and for our purpose gives a sufficiently good
approximation even for relatively light sparticles.%

\begin{table}[t] \centering
$%
\begin{tabular}
[c]{llll}\hline
LFV transitions: &  & Electric dipole moments: & \\\hline
$B\left(  \mu\rightarrow e\gamma\right)  <1.2\times10^{-11}$ &
$\text{\cite{Expmueg}}$ & $|d_{e}|<1.6\;10^{-27}\;e\,$cm &
$\text{\cite{ExpdTl}}$\\
$B\left(  \tau\rightarrow e\gamma\right)  <1.1\times10^{-7}$ &
$\text{\cite{Exptaueg}}$ & $|d_{\mu}|<1.8\;10^{-19}\;e\,$cm &
$\text{\cite{Expdmu}}$\\
$B\left(  \tau\rightarrow\mu\gamma\right)  <4.5\times10^{-8}$ &
$\text{\cite{Exptaumug}\ \ }$ & $d_{\tau}\in\lbrack-2.2,4.5]\times
10^{-17}\;e\,$cm & $\text{\cite{Expdtau}}$\\\hline
\end{tabular}
\ \ \ \ \ \ $%
\caption{Current experimental bounds on LFV transitions $\ell^I \rightarrow
\ell^J \gamma$
and leptonic EDMs. The electron, muon and tau EDMs are derived from the
bound on the thallium EDM, from the $(g-2)_\mu$ experiment, and from
the $e^+e^- \rightarrow\tau^+\tau^-$ data, respectively.}%
\label{TableExp}%
\end{table}%

The experimental situation is shown in Table~\ref{TableExp}. Several of these
measurements are expected to be improved in the near future, with
sensitivities reaching $10^{-29}$ $e\,$cm for $d_{e}$, $10^{-24}$ $e\,$cm for
$d_{\mu}$, $10^{-13}-10^{-14}$ for $B(\mu\rightarrow e\gamma)$ at the MEG
experiment\cite{MEG}, and $2\times10^{-9}$ for $B(\tau\rightarrow(e,\mu
)\gamma)$ at $B$ factories (for reviews, see e.g.
Refs.~\cite{PospelovR05,CERNreport}).

We now describe the input values for the various parameters needed to estimate
the EDMs and LFV transitions. All the parameters entering $\mathbf{Y}_{\nu
}^{\dagger}\mathbf{Y}_{\nu}$ and $\mathbf{Y}_{e}$ are fixed at the electroweak
scale. If one were to impose MFV at the high-energy scale $M_{R}$ or above,
RGE effects would play an important role~\cite{BorzumatiM86}. However, the MFV
expansions are RGE invariant, so all the RGE effects can be described through
running MFV coefficients. In the squark sector, it was shown in
Ref.~\cite{ParadisiRRS08,ColangeloNS08} that the coefficients at the low scale
are at most $\mathcal{O}(1)$ if they were of $\mathcal{O}(1)$ at the GUT
scale, and the same should hold also for the slepton sector (though the
converse is not necessarily true). Therefore, in the present work, to avoid
prescribing anything about the dynamics at the high-scale, we impose MFV at
the electroweak scale and allow for a rather generous range of variation for
the MFV coefficients (see Eq.~(\ref{MFVcoeffs}) below).

First, neutrino mixing parameters are taken from the best-fit of
Ref.~\cite{NeutrinoData}%
\begin{gather}
\Delta m_{21}^{2}=\Delta m_{\odot}^{2}=7.65_{-0.20}^{+0.23}\times
10^{-5}\,\text{eV}^{2},\;|\Delta m_{31}^{2}|=\Delta m_{atm}^{2}=2.4_{-0.11}%
^{+0.12}\times10^{-3}\,\text{eV}^{2}\;,\nonumber\\
\sin^{2}\theta_{\odot}=0.304_{-0.016}^{+0.022},\;\sin^{2}\theta_{atm}%
=0.50_{-0.06}^{+0.07},\;\sin^{2}\theta_{13}\leq0.056\;. \label{NData}%
\end{gather}
The neutrino mass-scale $m_{\nu}\equiv m_{\nu1}$ is unknown, but should not
exceed about $1$ eV if the cosmological bound $\sum_{i}m_{i}\lesssim1\;$eV
holds~\cite{Cosmo}. For $M_{R}$ we enforce the perturbativity condition
$|\mathbf{Y}_{\nu}^{\dagger}\mathbf{Y}_{\nu}|\lesssim1$, which translates as
$M_{R}\lesssim10^{13}$ GeV for $m_{\nu}\simeq1$\thinspace eV, and goes up to a
few $10^{14}$ GeV when $m_{\nu}\simeq0$. The $\phi_{i}$ parameters affect both
the phase and norm of $\mathbf{Y}_{\nu}^{\dagger}\mathbf{Y}_{\nu}$, and we
allow them to vary between $\pm1/2$ so that they never upset the
perturbativity bound (larger values are in any case disfavored by bounds on
LFV transitions, see next section). Finally, the other CP-violating parameters
are varied throughout their allowed ranges. In summary, we take%
\begin{equation}%
\begin{tabular}
[c]{lllll}%
$\gamma,\alpha,\beta$ & $\in\left[  -\pi,\,+\pi\right]  \;,$ &  & $m_{\nu}%
\;$(eV) & $\in\left[  0,\,0.1\right]  \;,$\\
$\sin^{2}\theta_{13}$ & $\in\left[  0,\,0.056\right]  \;,$ &  & $M_{R}%
\;$(GeV) & $\in\lbrack10^{9},\,10^{14}]\;,$\\
$\phi_{1},\phi_{2},\phi_{3}$ & $\in\left[  -1/2,\,+1/2\right]  \;,$ &  &  &
\end{tabular}
\ \ \ \ \ \label{ranges}%
\end{equation}
while the solar and atmospheric mass-differences and angles are fixed to their
central values in Eq.~(\ref{NData}).

For the MSSM parameters, we vary $\tan\beta\equiv v_{u}/v_{d}$ between $10$
and $50$, but fix the gaugino masses and the $\mu$ term to $\mu=\pm400$~GeV,
$M_{1}=200~$GeV and $M_{2}=400~$GeV at the electroweak scale. Since our goal
is to analyze the consequences of the phases in the slepton sector, we take
them real at that scale (so that RGE effects do not regenerate their phases
from the complex trilinear couplings~\cite{OlivePRS05}). With these
parameters, the tree-level chargino and neutralino masses are
\begin{subequations}
\label{Gaugino}%
\begin{align}
m_{\chi_{1}^{\pm}}  &  \approx354\text{ GeV},\;\;m_{\chi_{2}^{\pm}}%
\approx456\text{ GeV},\;\;\\
m_{\chi_{1}^{0}}  &  \approx198\text{ GeV},\;\;m_{\chi_{2}^{0}}\approx
354\text{ GeV},\;\;m_{\chi_{3}^{0}}\approx407\text{ GeV},\;\;m_{\chi_{4}^{0}%
}\approx455\text{ GeV}\;.
\end{align}
The slepton masses follow from the soft-breaking terms Eq.~(\ref{MFV}). We fix%
\end{subequations}
\begin{equation}
m_{0}=600\text{ GeV},\;\;A_{0}=400\text{ GeV }, \label{SUSYbreak}%
\end{equation}
and scan over%
\begin{equation}
a_{i},b_{i},\operatorname{Re}c_{i},\operatorname{Re}d_{i},\operatorname{Im}%
c_{i},\operatorname{Im}d_{i}\in\pm\;\left[  0.1,8\right]  \;,
\label{MFVcoeffs}%
\end{equation}
discarding points which lead to slepton or sneutrino masses below $100$ GeV.

\subsection{Leptonic EDMs versus LFV transitions}

\begin{figure}[pth]
\centering    \includegraphics[width=16.0cm]{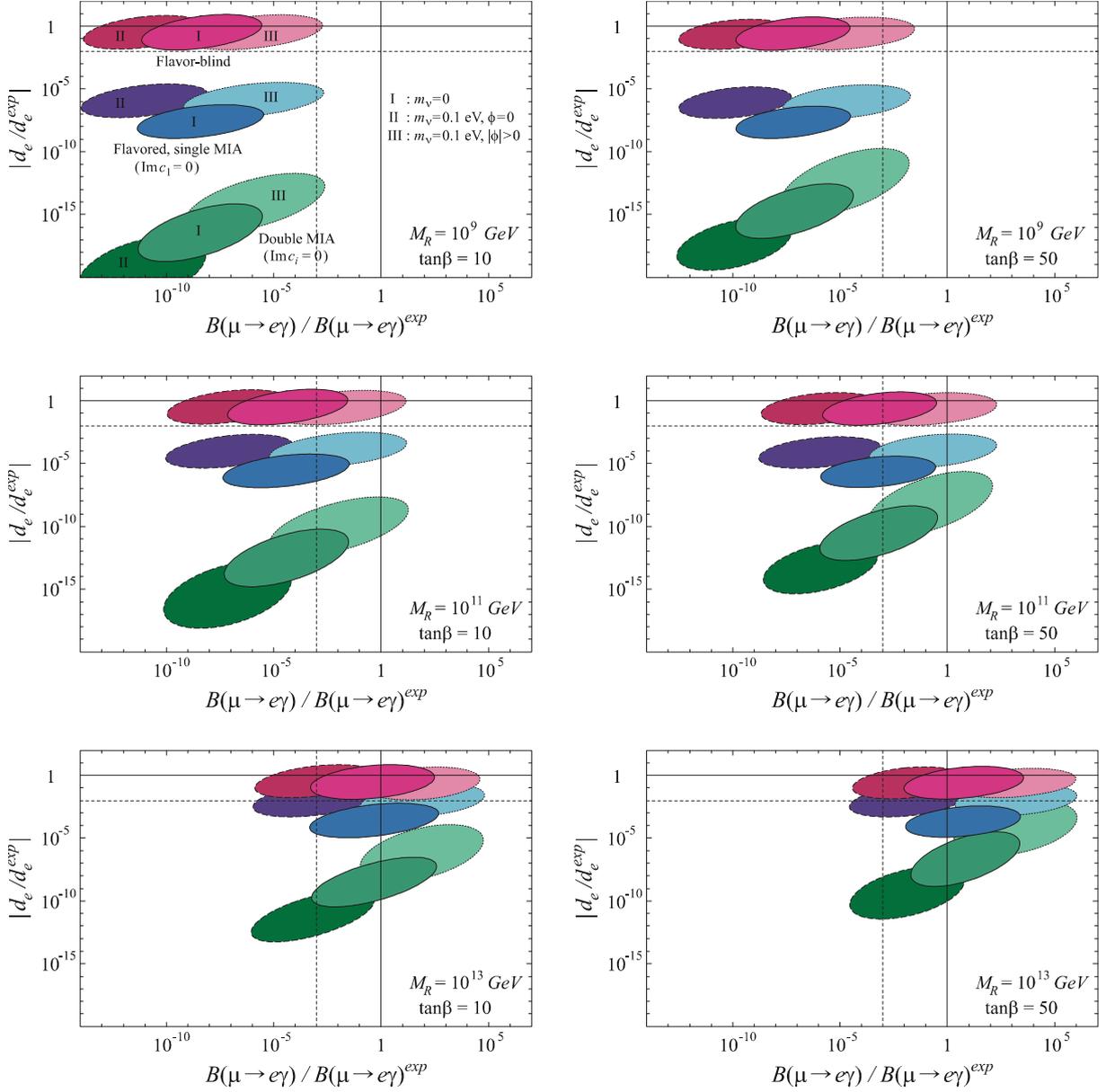}  \caption{Contours at
90\% in the $B\left(  \mu\rightarrow e\gamma\right)  -d_{e}$ plane (normalized
to their experimental bounds) corresponding to scanning over the parameters as
given in Eqs.~(\ref{ranges}) and (\ref{MFVcoeffs}), but subjected to the
constraints of the scenarios in Table~2 and Eq.~(\ref{scenars}). The contours
for $m_{\nu}=0$ are drawn for real spurions only. Those for complex spurions
mostly overlap and extend them by at most one order of magnitude towards
larger $B\left(  \mu\rightarrow e\gamma\right)  $ without any shift in $d_{e}%
$. Finally, the vertical and horizontal dashed lines show the expected
sensitivities of the next generation of experiments searching for $B\left(
\mu\rightarrow e\gamma\right)  $ and $d_{e}$, respectively.}%
\label{Fig1}%
\end{figure}

The numerical discussion is organized into several scenarios. The two spurions
$\mathbf{Y}_{e}^{\dagger}\mathbf{Y}_{e}$ and $\mathbf{Y}_{\nu}^{\dagger
}\mathbf{Y}_{\nu}$ vary according to different parameters: the former is tuned
entirely by $\tan\beta$, while the latter is linear in $M_{R}$, see
Eq.~(\ref{Spurion}). Hence, to span a large range of possibilities, we
consider separately%
\begin{equation}
M_{R}=10^{9},10^{11},10^{13}\,\text{GeV},\;\;\tan\beta=10,50\;.
\label{scenars}%
\end{equation}
Given the MSSM mass spectrum specified in Eqs.~(\ref{Gaugino}--\ref{MFVcoeffs}%
), larger values for $M_{R}$ are disfavored by the current bounds on LFV
transitions, as will be detailed below. For such values, the neutrino Yukawa
couplings $\mathbf{Y}_{\nu}$ are relatively small, powers of $\mathbf{Y}_{\nu
}^{\dagger}\mathbf{Y}_{\nu}$ become very suppressed, and the MFV expansions
simplify to (remember $\mathbf{A}\equiv\mathbf{Y}_{e}^{\dagger}\mathbf{Y}_{e}$
and $\mathbf{B}\equiv\mathbf{Y}_{\nu}^{\dagger}\mathbf{Y}_{\nu}$)
\begin{subequations}
\label{MFVred}%
\begin{align}
\mathbf{m}_{L}^{2}  &  =m_{0}^{2}(a_{1}\mathbf{1}+a_{2}\mathbf{A}%
+a_{3}\mathbf{B}+a_{5}\{\mathbf{A},\mathbf{B}\}+ib_{1}[\mathbf{A}%
,\mathbf{B}])\;,\\
\mathbf{m}_{E}^{2}  &  =m_{0}^{2}(a_{7}\mathbf{1}+\mathbf{Y}_{e}%
(a_{8}\mathbf{1}+a_{9}\mathbf{B}+a_{11}\{\mathbf{A},\mathbf{B}\}+ib_{4}%
[\mathbf{A},\mathbf{B}])\mathbf{Y}_{e}^{\dagger})\;,\\
\mathbf{A}_{e}  &  =A_{0}\mathbf{Y}_{e}(c_{1}\mathbf{1}+c_{2}\mathbf{A}%
+c_{3}\mathbf{B}+c_{5}\{\mathbf{A},\mathbf{B}\}+d_{1}i[\mathbf{A}%
,\mathbf{B}])\;.
\end{align}
Further, numerically, the operator $a_{5}$ and $a_{11}$ are at most of $30\%$
relative to the leading contributions and could be neglected in a first
approximation. Note that since $\mathbf{Y}_{e}^{\dagger}\mathbf{Y}_{e}$ is
diagonal, $\mathbf{Y}_{\nu}^{\dagger}\mathbf{Y}_{\nu}$ should not be discarded
everywhere since it is the only source of flavor transitions, no matter how
small is $M_{R}$. Similarly, at low $\tan\beta$, the suppressed $\mathbf{Y}%
_{e}^{\dagger}\mathbf{Y}_{e}$ is needed to account for CP-violation in the
$\mathbf{m}_{L}^{2}$ and $\mathbf{m}_{E}^{2}$ sectors (see the $b_{1}$ and
$b_{4}$ operators).%

\begin{table}[t] \centering
$%
\begin{tabular}
[c]{|l|lll|}\hline
Constraints on the MFV coefficients: & \multicolumn{3}{|c|}{Constraints on the
$\mathbf{Y}_{\nu}^{\dagger}\mathbf{Y}_{\nu}$ spurion:}\\
& $%
\begin{array}
[c]{c}%
m_{\nu}=0\\
\phi_{i}=0
\end{array}
$ & $%
\begin{array}
[c]{c}%
m_{\nu}=0.1\,eV\\
\phi_{i}=0
\end{array}
$ & $%
\begin{array}
[c]{c}%
m_{\nu}=0.1\,eV\\
\phi_{i}\neq0
\end{array}
$\\\cline{2-4}%
All possible complex coefficients & Red & Dark red & Light red\\
No flavor-blind phase ($\operatorname{Im}c_{1}=0$) & Blue & Dark blue & Light
blue\\
No flavor-diagonal phases ($\operatorname{Im}c_{i}=0$) & Green & Dark green &
Light green\\\hline
\end{tabular}
\ \ \ \ \ \ $%
\caption{Scenarios for the CP-violating phases coming from the spurion
$\mathbf{Y}_{\nu}^{\dagger}\mathbf{Y}_{\nu}%
$ and from the MFV coefficients. It is
understood that $\phi
_i$ denotes collectively all the nine CP-violating phase entering
$\mathbf{Y}_{\nu}$, i.e. the Dirac phase $\gamma$, the two Majorana phases
$\alpha$ and $\beta$, and the three $\phi_i$ parameters are varied within the
ranges (\ref{ranges}%
). The colors correspond to those of the 90\% contours in Fig.1.
The situation with $m_\nu= 0$, $\phi_i \neq
0$ is essentially identical to $m_\nu= 0$,
$\phi
_i = 0$. Finally, it is understood that the normal spectrum is supposed for neutrino
masses.}
\end{table}%

To analyze the impact of the different types of CP-violating phases on the
EDMs, we distinguish nine scenarios, each specified by a set of conditions
imposed on the MFV coefficients on one hand, and on the parameters entering
the spurion $\mathbf{Y}_{\nu}^{\dagger}\mathbf{Y}_{\nu}$ on the other, as
summarized in Table~2. For each of these nine scenarios, the results of
scanning over the parameters according to flat distributions (and subject to
the corresponding constraints) are shown in Fig.~\ref{Fig1} as 90\% contours
in the $B\left(  \mu\rightarrow e\gamma\right)  -d_{e}$ plane (normalized to
their respective experimental bound). We do not give any statistical
interpretation to these contours: one can understand the 90\% limit as a
simple procedure to remove peculiar, fine-tuned situations in parameter space.
Finally, we also discard points for which $a_{\mu}$ is not acceptable. Given
the current discrepancy between theory and experiment of about $\left(
30\pm10\right)  \times10^{-10}$ (see Ref.~\cite{CERNreport} and references
therein), we keep points for which
\end{subequations}
\begin{equation}
0\leq a_{\mu}^{\text{SUSY}}\leq40\times10^{-10}\;.
\end{equation}
Apart from fixing the sign of $\mu$, given our mass spectrum, only less than
5\% of the points have to be thrown away, and this only for large $\tan\beta$.

\subsubsection{LFV transitions}

As visible in Fig.~1, the LFV transitions are insensitive to the constraints
imposed on the CP-violating part of the MFV coefficients, in stark contrast to
the EDMs, as will be discussed below. This can be understood from the fact
that the amplitudes for the LFV processes are always dominated by the
$\delta_{LL}^{JI}$ term in Eq.~(\ref{LFV}). Given the scaling of the operators
in Eq.~(\ref{MFVred}), we can write Eq.~(\ref{LFV}) as:%
\begin{equation}
B\left(  \ell^{I}\rightarrow\ell^{J}\gamma\right)  \sim\frac{M_{W}^{4}%
\,M_{1}^{2}\tan^{2}\beta}{|\mu|^{2}}\left|  \delta_{LL}^{JI}\mathcal{F}%
_{1}\right|  ^{2}\;,\;\;\delta_{LL}\approx\frac{a_{3}}{a_{1}}\mathbf{Y}_{\nu
}^{\dagger}\mathbf{Y}_{\nu}+\frac{a_{5}}{a_{1}}\{\mathbf{Y}_{e}^{\dagger
}\mathbf{Y}_{e},\mathbf{Y}_{\nu}^{\dagger}\mathbf{Y}_{\nu}\}\;, \label{LFV2}%
\end{equation}
with, for the $\mu\rightarrow e\gamma$ transition, $\delta_{LL}^{12}$ fully
dominated by the $a_{3}$ operator (see Fig.~2A)%
\begin{equation}
\delta_{LL}^{12}\approx\frac{a_{3}}{a_{1}}(\mathbf{Y}_{\nu}^{\dagger
}\mathbf{Y}_{\nu})^{12}=\frac{a_{3}}{a_{1}}\frac{M_{R}}{v_{u}^{2}%
}(U(\mathbf{m}_{\nu}^{1/2})\,e^{2i\mathbf{\Phi}}\mathbf{\,}(\mathbf{m}_{\nu
}^{1/2})U^{\dagger})^{12}\;. \label{deltaLL}%
\end{equation}
In these expressions, the operator $b_{1}$ is absent: it is competitive only
for the imaginary parts of the off-diagonal entries of $\delta_{LL}^{JI}$, not
for their norms. The loop function $\mathcal{F}_{1}$ has a strong
non-polynomial dependence on the masses, hence also on $a_{1}$ since
$m_{L}^{2}\approx m_{0}^{2}a_{1}$ when $M_{R}\lesssim10^{13}$ GeV and
$\tan\beta$ is not too large. Nevertheless, this dependence is monotonic, and
Eq.~(\ref{deltaLL}) is sufficient to understand the behavior of $B\left(
\mu\rightarrow e\gamma\right)  $ as the MFV and neutrino parameters are varied.

\paragraph{\textbf{Sparticle mass dependences:}}

The LFV decay rates are quadratic in $M_{R}$, and increasing this parameter
beyond $10^{13}$ GeV would violate the experimental bound for most values of
$a_{3}$ and $a_{1}$. This maximal value for $M_{R}$ follows from the choice we
made for the MSSM mass spectrum, Eqs.~(\ref{Gaugino}) and (\ref{SUSYbreak}).
For instance, taking $m_{0}$ larger so as to make sleptons and sneutrinos
heavier would shift the center of the contours towards the lower left corner,
suppressing both EDMs and LFV transitions, and thus allowing for slightly
larger $M_{R}$. Still, note that with $m_{0}=600$ GeV and $a_{1}$ between
$0.1$ and $8$, we are already scanning over a large range of slepton masses as
$m_{L}$ varies between about $100$ GeV and $2$ TeV. Further, the contours in
Fig.~1 span several orders of magnitude for $B\left(  \mu\rightarrow
e\gamma\right)  $, but the corresponding $d_{e}$ values are much more
concentrated. Therefore, changing $m_{0}$ affects more $B\left(
\mu\rightarrow e\gamma\right)  $ than $d_{e}$, and $m_{0}=600$ GeV appears as
a reasonably small value still allowing to probe relatively large $M_{R}$
without violating the current $B\left(  \mu\rightarrow e\gamma\right)  $ bound.

\begin{figure}[pt]
\centering                          \includegraphics[width=16.0cm]{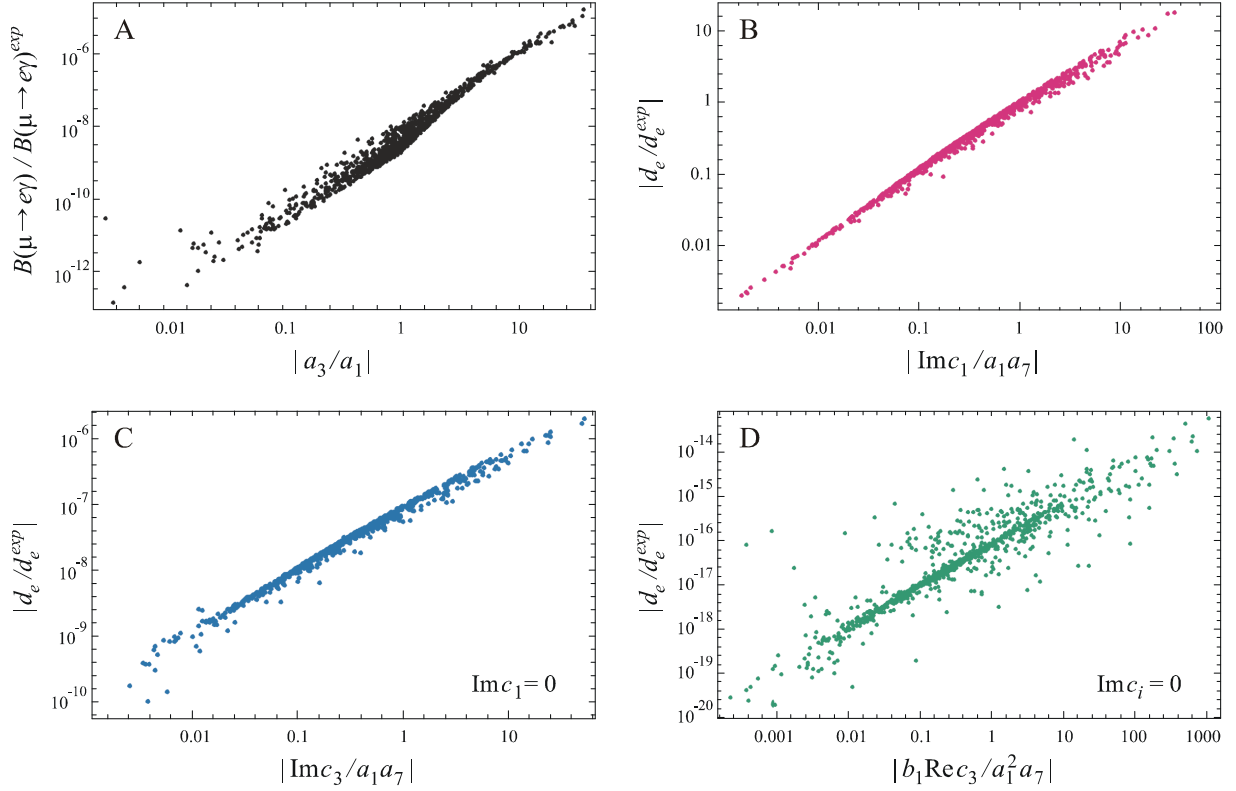}
\caption{Illustration of the hierarchical dominance of a single phase per
scenario for $M_{R}=10^{9}$ GeV, $\tan\beta=10$, and $m_{\nu}=0$. Colors refer
to Fig.~1. A.~Dominance of the $\delta_{LL}^{12}\approx a_{3}/a_{1}$
contribution for the $B\left(  \mu\rightarrow e\gamma\right)  $ transition.
The small spread of the points is due to the dependences of the loop functions
on the masses, especially strong for light sleptons ($\approx$ small $a_{1}$).
B.~Dominance of the flavor-blind phase $\operatorname{Im}c_{1}$ for $d_{e}$.
Again, the small spread of the points is due to the loop function.
C.~Dominance of the subleading flavor-diagonal phase $\operatorname{Im}c_{3}$.
D.~Approximate dominance of the flavor off-diagonal phase from $b_{1}$. Here,
the spread of points is also due to additional operators whose contributions
can be competitive, especially for light sparticles (small $a_{1}$ and/or
$a_{7}$) or when $b_{1}$ and/or $\operatorname{Re}c_{3}$ are small. The effect
of taking $\tan\beta$ larger affects C and D, taking $M_{R}$ larger affects D,
while taking $m_{\nu}$ larger and including fixed CP-violating phases in the
spurion does not affect the dominance, but shifts the points to smaller or
larger values, as shown in Fig.~1.}%
\label{Fig2}%
\end{figure}

\paragraph{Neutrino spurion parameters:}

When the lightest neutrino is massless (in the normal spectrum), the
$\mu\rightarrow e\gamma$ transition is essentially independent of the
CP-violating phases in the spurion $\mathbf{Y}_{\nu}^{\dagger}\mathbf{Y}_{\nu
}$ (given their range of variation specified in (\ref{ranges})). When $m_{\nu
}$ increases, either $B\left(  \mu\rightarrow e\gamma\right)  $ is suppressed
if $\phi_{i}=0$ or increased if the $\phi_{i}$ are varied in their ranges.
This is easy to understand from the behavior of the off-diagonal entries of
$\mathbf{Y}_{\nu}^{\dagger}\mathbf{Y}_{\nu}$. When $\phi_{i}=0$, they are
tuned entirely by the neutrino mass-differences $m_{\nu3}-m_{\nu}$ or
$m_{\nu2}-m_{\nu}$, which decrease with increasing $m_{\nu}$ keeping $\Delta
m_{\odot}^{2}$ and $\Delta m_{atm}^{2}$ fixed to their experimental values. On
the contrary, when $\phi_{i}\neq0$, off-diagonal entries receive a
contribution linear in $m_{\nu}$ and $\phi_{i}$ (see Eq.~(\ref{Spurion})),%
\begin{equation}
(\mathbf{Y}_{\nu}^{\dagger}\mathbf{Y}_{\nu})^{12}=\frac{M_{R}}{\sqrt{2}%
v_{u}^{2}}\left(  s_{\odot}c_{\odot}\Delta m_{21}+2im_{\nu}\left(  \phi
_{1}+c_{\odot}\phi_{2}+s_{\odot}\phi_{3}\right)  +...\right)  \;,
\label{offspurions}%
\end{equation}
which dominates for $m_{\nu}>\Delta m_{21}$, given the ranges (\ref{ranges})
for the $\phi_{i}$ parameters.

\paragraph{Correlations and tan$\beta$:}

As long as there is only the $a_{3}$ operator, the three LFV transitions
$\mu\rightarrow e\gamma$, $\tau\rightarrow\mu\gamma$ and $\tau\rightarrow
e\gamma$ are clearly correlated among themselves. The ratios of their width
can be predicted in terms of the spurion parameters only, i.e. in terms of the
relevant off-diagonal entries of $\mathbf{Y}_{\nu}^{\dagger}\mathbf{Y}_{\nu}$,
and are thus identical to those already extensively studied in the context of
the supersymmetric seesaw. For example, these ratios show a strong sensitivity
to the $\phi_{i}$ parameters and a further dependence on the Majorana
phases~\cite{CasasI01,PascoliPY03,PetcovS,BrancoEtAl06}. Note though that this
sensitivity to neutrino parameters is smaller for the $\tau\rightarrow
\mu\gamma$ mode, tuned by the larger $\Delta m_{atm}^{2}$ in the limit
$m_{\nu}=0$.

At large $\tan\beta$, the situation changes slightly because the operator
$a_{5}$ becomes competitive for $\tau\rightarrow\mu\gamma$ and $\tau
\rightarrow e\gamma$. Though the ratio of the widths of these two modes still
depends only on the spurion parameters, their ratios to $B(\mu\rightarrow
e\gamma)$ are function also of the MFV parameters $a_{5}$ and $a_{3}$.
However, the operator $a_{5}$ never becomes dominant, and thus never changes
the order of magnitude of the $\tau\rightarrow\mu\gamma$ and $\tau\rightarrow
e\gamma$ modes.

Altogether, the predictions for the $\tau\rightarrow\mu\gamma$ or
$\tau\rightarrow e\gamma$ modes within MFV are always at least two orders of
magnitude farther away from their respective experimental limits than
$\mu\rightarrow e\gamma$. Therefore, in the MFV\ setting, $\mu\rightarrow
e\gamma$ currently gives the best constraints, or opportunity for discovery.
In Fig.~1 we have drawn the reach of the MEG experiment, which will probe this
mode down to the $10^{-13}-10^{-14}$ range, and will essentially exclude
$M_{R}\gtrsim10^{13}$ GeV with our range of sparticle masses.

\subsubsection{Leptonic EDMs}

Let us now turn to the lepton EDMs, starting from Eq.~(\ref{EDM}). As will be
explored in detail below, only three contributions are actually relevant:%
\begin{equation}
\frac{d_{I}}{e}\approx\frac{-m_{e^{I}}\,\alpha M_{1}A_{0}}{8\pi\cos^{2}%
\theta_{W}m_{0}^{4}}\left(  \frac{\operatorname{Im}c_{1}}{a_{1}a_{7}%
}+\frac{\operatorname{Im}c_{2}\mathbf{Y}_{e}^{\dagger}\mathbf{Y}%
_{e}+\operatorname{Im}c_{3}\mathbf{Y}_{\nu}^{\dagger}\mathbf{Y}_{\nu}}%
{a_{1}a_{7}}-\frac{b_{1}\operatorname{Re}c_{3}}{a_{1}^{2}a_{7}}[\mathbf{Y}%
_{e}^{\dagger}\mathbf{Y}_{e},\mathbf{Y}_{\nu}^{\dagger}\mathbf{Y}_{\nu
}]\mathbf{Y}_{\nu}^{\dagger}\mathbf{Y}_{\nu}+...\right)  ^{II}\;. \label{EDMe}%
\end{equation}
The first two terms come from the $-A_{e}^{I\ast}$ piece in Eq.~(\ref{EDM}),
i.e. the diagonal entries of the trilinear term $\mathbf{A}_{e}$, while the
last originates from the double mass-insertion $\delta_{LL}^{IK}\delta
_{LR}^{KI}$, i.e. from $\mathbf{m}_{L}^{2}$ and $\mathbf{A}_{e}$. Contrary to
the case of LFV transitions, the loop functions $\mathcal{F}_{5}$ and
$\mathcal{F}_{8}$ are much flatter for our range of masses, and can be
approximated by $|\mu^{2}|/m_{L}^{2}m_{R}^{2}\approx|\mu^{2}|/(m_{0}^{4}%
a_{1}a_{7})$.

To show the main dependences on the neutrino parameters, let us write $d_{e}$
in the $m_{\nu}=0$ limit, keeping $\mathbf{Y}_{\nu}^{\dagger}\mathbf{Y}_{\nu}$
real:%
\begin{equation}
\frac{d_{e}}{e}\approx\frac{-m_{e}\,\alpha M_{1}}{8\pi\cos^{2}\theta_{W}%
}\frac{A_{0}}{m_{0}^{4}}\left(  \frac{\operatorname{Im}c_{1}}{a_{1}a_{7}%
}+\frac{M_{R}\Delta m_{21}s_{\odot}^{2}}{v_{u}^{2}}\frac{\operatorname{Im}%
c_{3}}{a_{1}a_{7}}-\frac{M_{R}^{2}c_{\odot}^{2}s_{\odot}^{2}\left(  \Delta
m_{21}\right)  ^{2}m_{\tau}^{2}}{2v_{d}^{2}v_{u}^{4}}\frac{b_{1}%
\operatorname{Re}c_{3}}{a_{1}^{2}a_{7}}+...\right)  \;. \label{Hierach}%
\end{equation}
The term $\operatorname{Im}c_{2}$ does not contribute: being quadratic in the
lepton mass, it is relevant only for $d_{\mu}$ and $d_{\tau}$. The striking
feature of $d_{e}$ within MFV is the very strong hierarchy between these three
terms. Each of them corresponds to one scenario for the MFV coefficients given
in Table~2, and their strong hierarchy is obvious in Fig.~1. Analytically,
Eq.~(\ref{Hierach}) shows that the contribution from the flavor-blind
$\operatorname{Im}c_{1}$ is much larger than the one from the flavor-diagonal
phase $\operatorname{Im}c_{3}$, linear in $\Delta m_{21}$, which is itself
much larger than the double MIA contribution of the flavor off-diagonal phase
$b_{1}$, quadratic in $\Delta m_{21}$, but this hierarchy gets mitigated as
$M_{R}$ increases. Also, only the third term shows a strong, quadratic
sensitivity to $\tan\beta$. All this is immediately apparent in Fig.~1.

Let us look more closely at each scenario.

\paragraph{The dominant flavor-blind phase:}

When $c_{1}$ is complex, it completely dominates, as seen in Fig.~1 and
Fig.~2B. This scenario corresponds essentially to what has been analyzed in
Ref.~\cite{AltmannshoferBP08} (see also Ref.~\cite{EllisLP08}). Contrary to
the LFV transitions, which always scale at least quadratically with $M_{R}$,
this dominant contribution is independent of $M_{R}$. Further, it is clearly
independent of the neutrino parameters. The important point is that even for
the relatively light MSSM mass-spectrum specified in Eqs.~(\ref{Gaugino}%
--\ref{MFVcoeffs}), $d_{e}$ can easily satisfy its experimental upper bound.
For example, with $\operatorname{Im}c_{1}=1$, one immediately reads from
Fig.~2B that the current bound on $d_{e}$ imposes $a_{1}a_{7}\gtrsim1$, which
corresponds to $m_{L}^{2}\approx m_{R}^{2}\gtrsim600$ GeV when $a_{1}=a_{7}$.
On the contrary, the expected two order-of-magnitude improvement in the
measurement of $d_{e}$ would rule out this scenario, except for prohibitively
high slepton masses, or unnaturally small $\operatorname{Im}c_{1}$. The same
conclusion was reached in Ref.~\cite{AltmannshoferBP08}. The lower bound
$d_{e}\gtrsim10^{-28}$ $e\,$cm they found is comparable (though slightly
larger) than the one we can extract from Fig.~1.

At moderate $\tan\beta$, only $c_{1}$ contributes and the relative sizes of
$d_{e}$, $d_{\mu}$ and $d_{\tau}$ are simply ruled by the ratios of the lepton
masses. For our specific MSSM mass spectrum, this translates as%
\begin{equation}
\frac{d_{\mu}}{d_{\mu}^{\exp}}\approx\frac{m_{\mu}}{m_{e}}\frac{d_{e}}{d_{\mu
}^{\exp}}\lesssim10^{-5},\;\;\frac{d_{\tau}}{d_{\tau}^{\exp}}\approx
\frac{m_{\tau}}{m_{e}}\frac{d_{e}}{d_{\tau}^{\exp}}\lesssim10^{-6}\;.
\label{RatioEDM}%
\end{equation}
This means that while $d_{e}$ is already around its current experimental
bound, future experiments aiming at $d_{\mu}$ and $d_{\tau}$ should gain
respectively five and six orders of magnitude in sensitivity to be just barely
competitive. This conclusion is unchanged at large $\tan\beta$, because even
though the MFV\ operator $c_{2}$ contributes for $d_{\tau}$, it always affects
the scaling (\ref{RatioEDM}) by less than an order of magnitude.

\paragraph{The subdominant flavor diagonal phases:}

If the flavor-blind phase is turned off by setting $\operatorname{Im}c_{1}=0$,
the electron EDM is generated entirely by $\operatorname{Im}c_{3}$, see
Fig.~2C. The dependence on the neutrino parameters is now apparent in
Fig.~\ref{Fig1}: while $d_{e}$ is essentially independent of the $\phi_{i}$'s
no matter $m_{\nu}$, it increases with $m_{\nu}$ as can be understood looking
at the diagonal entries of $\mathbf{Y}_{\nu}^{\dagger}\mathbf{Y}_{\nu}$. In
any case, except for very large $M_{R}$, $d_{e}$ is rather far from its
experimental bound and the additional CP-violating phases in the trilinear
sector brought in by MFV are completely free. When $M_{R}$ becomes very large,
except for some contrived set of spurion parameters, the LFV transitions fail
to pass their experimental bounds well before $d_{e}$.

In this scenario, the EDMs no longer scale as the lepton masses. Even if there
were only the $\operatorname{Im}c_{3}$ contribution, the ratios of EDMs would
depend on the neutrino parameters. For example, $d_{\mu}$ and $d_{\tau}$ would
be enhanced by $\Delta m_{atm}/\Delta m_{sol}$ when $m_{\nu}=0$. But in
addition to $\operatorname{Im}c_{3}$, which always dominates for $d_{e}$, the
$\operatorname{Im}c_{2}$ contribution is competitive for $d_{\mu}$ when
$M_{R}\lesssim10^{11}$, and dominates $d_{\tau}$ for $M_{R}\lesssim10^{13}$
GeV, even at low $\tan\beta$. Further, for $\tan\beta=50$, the tau Yukawa
coupling is of $\mathcal{O}(1)$, $m_{\tau}^{2}/v_{d}^{2}\approx1/4$, bringing
$d_{\tau}$ very close to the values attainable with the flavor-blind phase.
This makes $d_{\tau}$ particularly interesting to test the presence of new
CP-violating phases in the slepton sector, both flavor-blind and flavor
diagonal. However, the bound (\ref{RatioEDM}) cannot be evaded, and $d_{\tau}$
should remain beyond the experimental reach for the near future.

\paragraph{The tiny flavor off-diagonal phases:}

When all the flavor-diagonal phases are turned off by setting
$\operatorname{Im}c_{i}=0$, the EDMs are generated by second order effects in
the mass insertions. As explained in Section 2, this is the order at which we
would expect to see a strong sensitivity to the CP-violating phase in the
$\mathbf{Y}_{\nu}^{\dagger}\mathbf{Y}_{\nu}$ spurion (the Dirac and Majorana
phases and $\phi_{i}$ parameters), but these turn out to be subleading. The
dominant CP-violating effect when $\operatorname{Im}c_{i}=0$ comes instead
from the CP-violating $b_{1}$ parameters taken together with
$\operatorname{Re}c_{3}$. Of course, this $b_{1}\operatorname{Re}c_{3}$
contribution shows a strong dependence on the $\phi_{i}$ when $m_{\nu}\neq0$
but this comes entirely from the sensitivity of the off-diagonal elements of
$\mathbf{Y}_{\nu}^{\dagger}\mathbf{Y}_{\nu}$ on the $\phi_{i}$, as can be seen
in Eq.~(\ref{offspurions}), and thus is not a CP-violating effect.\footnote{We
saw in Sec. 2 that $\operatorname{Im}c_{1}$ is never exactly zero, but is at
least as large as the Jarlskog invariant (\ref{Jarlskog}). We checked that the
contributions from $\operatorname{Im}c_{1}\sim J$ are always at least one
order of magnitude smaller than those from $b_{1}\operatorname{Re}c_{3}$.}

It should be stressed that the dominance of the $b_{1}\operatorname{Re}c_{3}$
contribution is not as strong as in the previous two scenarios, see Fig.~2D.
Other effects compete at the double MIA order, especially when $M_{R}%
\gtrsim10^{11}$ GeV or when $\tan\beta$ is large. For example, in the latter
case, the double MIA contribution to $d_{e}$ and $d_{\mu}$ comes from the
$\delta_{LL}^{I3}\delta_{LR}^{3I}$ term in Eq.~(\ref{EDM}), which involves the
third generation. As a result, the presence of additional operators in the
large $\tan\beta$ limit blurs the dominance of $b_{1}\operatorname{Re}c_{3}$.
On the other hand, $d_{\tau}$ is still dominated by $b_{1}\operatorname{Re}%
c_{3}$ since $\delta_{LR}^{33}=0$. This shows that though the dominance is not
complete, the overall picture can still be grasped with the help of the
$b_{1}\operatorname{Re}c_{3}$ contribution alone. In particular, whether these
competing effects are turned on or off does not visibly change the contours in Fig.~1.

In view of the previous comment, it is clear that the ratios of $d_{e}$,
$d_{\mu}$ and $d_{\tau}$ do not scale like the lepton masses. When all of them
are dominated by the single $b_{1}\operatorname{Re}c_{3}$ operator
combination, which requires low $M_{R}$ and $\tan\beta$, their ratios can be
predicted entirely in terms of the neutrino parameters. Still, being quadratic
in the off-diagonal entries of $\mathbf{Y}_{\nu}^{\dagger}\mathbf{Y}_{\nu}$,
these ratios are even more sensitive to these parameters than the ratios of
LFV transitions discussed before, and span several orders of magnitude when
$m_{\nu}>0$ and the $\phi_{i}$ vary in the ranges (\ref{ranges}). When $M_{R}$
or $\tan\beta$ are larger, additional operators enter and ratios of EDMs are
no longer fixed in terms of neutrino parameters.

Overall, the EDMs are very suppressed in the absence of CP-violating phases in
the $c_{i}$ operators, no matter the CP-violating phases in the spurions, see
Fig.~1. Only in the limit of very large $M_{R}$ is this suppression
compensated, but this is forbidden as LFV transitions violate their
experimental bounds much before the EDMs become even reasonably close to their
present limits.

\subsection{Comparison with the model-independent approach}

When MFV\ is used to directly parametrize the New Physics operator responsible
for $\ell^{I}\rightarrow\ell^{J}\gamma$ transitions, EDMs and MDMs, one
obtains~\cite{DambrosioGIS02,CiriglianoGIW05}%
\begin{align}
H_{eff}  &  =e\frac{1}{\Lambda^{2}}\frac{\alpha}{4\pi}\left(  \mathbf{Y}%
_{e}^{II}\mathcal{C}^{IJ}\right)  \bar{\psi}_{R}^{I}\sigma_{\mu\nu}\psi
_{L}^{J}F^{\mu\nu}H_{d}+h.c.\;,\nonumber\\
\mathcal{C}  &  =h_{1}\mathbf{1}+h_{2}\mathbf{A}+h_{3}\mathbf{B}%
+h_{4}\mathbf{B}^{2}+h_{5}\{\mathbf{A},\mathbf{B}\}+h_{6}\mathbf{BAB}%
\nonumber\\
&  \;\;\;\;\;\;\;+g_{1}i[\mathbf{A},\mathbf{B}]+g_{2}i[\mathbf{A}%
,\mathbf{B}^{2}]+g_{3}i(\mathbf{B}^{2}\mathbf{AB}-\mathbf{BAB}^{2})\;,
\label{EffOp}%
\end{align}
where $\mathcal{C}$ transforms as $\mathbf{Q}$ in Eq.~(\ref{HermQ}) and, as
before, $\mathbf{A}\equiv\mathbf{Y}_{e}^{\dagger}\mathbf{Y}_{e}$ and
$\mathbf{B}\equiv\mathbf{Y}_{\nu}^{\dagger}\mathbf{Y}_{\nu}$ with
$\mathbf{Y}_{e}$ diagonal in our basis. This expression is manifestly
invariant under the flavor group $U(3)^{5}$. As explained in Section 2, the
$U(3)^{5}$ symmetry does not force the coefficients to be real, hence we allow
all the $h_{i}$ and $g_{i}$ to be complex. The classification of the
CP-violating phases performed in Sec. 2.4 still holds: $\operatorname{Im}%
h_{1}$ is flavor-blind, $\operatorname{Im}h_{2-6}$ and $\operatorname{Im}%
g_{3}$ are flavor-diagonal, and $\operatorname{Im}g_{1,2}$ as well as all the
CP-violating phases in $\mathbf{Y}_{\nu}^{\dagger}\mathbf{Y}_{\nu}$ are flavor
off-diagonal since all the operators are hermitian. Note that the denomination
`flavor-blind' is to be understood as in the MSSM: strictly speaking, both the
$c_{1}$ operator in Eq.~(\ref{MFV}) and the $h_{1}$ operator in
Eq.~(\ref{EffOp}) are forbidden when $\mathbf{Y}_{e}$ is absent since the
whole $\mathbf{A}_{e}$ and $H_{eff}$ would be forbidden.

Though the MFV parametrization of the effective operator in Eq.~(\ref{EffOp})
is intended to be as model-independent as possible, we have introduced the
loop factor and gauge coupling explicitly (compare with Eq.~(\ref{ModIndep})).
Indeed, LFV and EDMs arise at the loop level in the MSSM, so it makes sense to
factor out these effects in the following comparison. From Eq.~(\ref{EffOp}),
the leptonic observables are (neglecting terms suppressed by $m_{e^{J}%
}/m_{e^{I}}$)%
\begin{equation}
\frac{B\left(  \ell^{I}\rightarrow\ell^{J}\gamma\right)  }{B\left(  \ell
^{I}\rightarrow\ell^{J}\bar{\nu}^{J}\nu^{I}\right)  }=\frac{\alpha\sin
^{4}\theta_{W}}{24\pi}\frac{M_{W}^{4}}{\Lambda^{4}}|\mathcal{C}^{IJ}%
|^{2},\;\;a_{I}=\frac{\alpha m_{e^{I}}^{2}}{\pi\Lambda^{2}}\operatorname{Re}%
\mathcal{C}^{II},\;\;\frac{d_{I}}{e}=\frac{\alpha m_{e^{I}}}{2\pi\Lambda^{2}%
}\operatorname{Im}\mathcal{C}^{II}\;. \label{EffObs}%
\end{equation}
The similarity with Eqs.~(\ref{LFV}) and (\ref{EDM}) is manifest in the limit
where all sparticles are degenerate (remember that the loop functions scale as
$\mathcal{F}_{i}\sim1/M_{\text{SUSY}}^{2}$). Let us compare the two MFV
implementations in detail.

\begin{itemize}
\item[--] The LFV transitions behave essentially as in the MSSM case studied
before. Indeed, the operator $h_{3}$ dominates, hence LFV transitions scale as
the off-diagonal entries of $\mathbf{Y}_{\nu}^{\dagger}\mathbf{Y}_{\nu}$,
exactly like in Eq.~(\ref{LFV2}). Correlations between LFV transitions are
thus similar to those for the supersymmetric
case~\cite{CiriglianoGIW05,CiriglianoG06,IsidoriMPT07}. The order of magnitude
is slightly smaller in the effective theory when $\Lambda\sim M_{\text{SUSY}}%
$, because of the different numerical factors and lack of the $\tan^{2}\beta$ enhancement.

\item[--] For the EDMs, the flavor blind phase $\operatorname{Im}h_{1}$
dominates. Comparing Eq.~(\ref{EffOp}) with (\ref{EDM}), $\operatorname{Im}%
h_{1}$ is the effective coupling describing both the effects of the
flavor-blind phase of $\mathbf{A}_{e}$ and of all the phases of the
flavor-blind parameters, including $\mu$. As a result, the matching with the
MSSM is slightly ambiguous. Indeed, taken at face-value, Eq.~(\ref{EffObs})
implies
\begin{equation}
\frac{\alpha m_{e^{I}}}{2\pi\Lambda^{2}}\lesssim1.6\times10^{-27}%
\;e\,\text{cm}\Rightarrow\Lambda\gtrsim2.7\;\text{TeV\ ,}%
\end{equation}
when $\operatorname{Im}h_{1}\sim\mathcal{O}(1)$. Compared to the MSSM, this is
an intermediate situation. An arbitrary phase for $\mu$ pushes $\Lambda$ well
above $5$ TeV, essentially because of the $\tan\beta$ enhancement. On the
contrary, the contribution of the flavor-blind phase of $\mathbf{A}_{e}$ is
easily suppressed by more than a factor of ten, once the precise mass
dependences are taken into account, pulling $\Lambda$ down below $800$ GeV
(see Fig.~2B).

\item[--] Concerning the flavor-diagonal phases, these are dominated by
$\operatorname{Im}h_{2}$ and $\operatorname{Im}h_{3}$, which are analogous to
the $\operatorname{Im}c_{2}$ and $\operatorname{Im}c_{3}$ contributions in the
MSSM, respectively. For $d_{e}$, only $\operatorname{Im}h_{3}$ contributes,
bringing in a suppression factor $M_{R}\Delta m_{21}/v_{u}^{2}$ (see
Eq.~(\ref{Hierach})). Interestingly, the model-independent formalism imposes a
(loose) correlation between LFV transitions and $d_{e}$, since they are all
due to the same MFV operator $h_{3}$. In the MSSM, this correlation is
completely absent because the LFV transitions are tuned by the slepton mass
term $\mathbf{m}_{L}^{2}$, while EDM by the trilinear coupling $\mathbf{A}%
_{e}$. For $d_{\mu}$ and $d_{\tau}$, both $\operatorname{Im}h_{2}$ and
$\operatorname{Im}h_{3}$ can contribute, exactly like in the MSSM. Finally, as
for the flavor-blind phase $\operatorname{Im}h_{1}$, significant numerical
factors affect these contributions and prevent a precise comparison of the
scale $\Lambda$ with the SUSY scale.

\item[--] The effects of the off-diagonal CP-violating phases are absent,
since there is no such thing as a double mass-insertion in the effective
operator formalism. Therefore, when $\operatorname{Im}h_{i}=0$, only the very
suppressed contribution of the flavor-diagonal phase of the $g_{3}$ operator
generates the EDMs. Though the suppression is less strong in the MSSM, the
situation is similar to the scenario with $\operatorname{Im}c_{i}=0$, see
Fig.~1. In other words, the effective operator formalism correctly predicts
that in the absence of flavor blind or diagonal phases, the EDMs are far too
suppressed to be seen in the near future.
\end{itemize}

Overall, the effective operator formalism catches the MSSM features quite
satisfactorily, provided one allows for some latitude in the numerical
coefficients. The loop factor and the gauge couplings obviously have to be
added by hand, but an additional factor of about ten should also be allowed
when extracting model-independent bounds on the scale $\Lambda$. This near
correspondence between the effective formalism and the MSSM holds even if the
latter has much more degrees of freedom, including more MFV coefficients,
because numerically, only a few operators are dominant in both cases, and they
have the same spurion structures. Still, some correlations between LFV
transitions and EDMs are absent in the MSSM, first because the EDMs depend on
both the left and right slepton masses while LFV transitions care only about
those of the left sleptons (see the occurrence of $a_{7}$ in Eq.~(\ref{EDM})),
and then simply because in the MSSM, different MFV coefficients enter in
$\delta_{LL}$ and $\delta_{LR}$.

\section{Beyond MFV: leptonic observables in the general MSSM}

In the general MSSM, the bounds on LFV processes and lepton EDMs are usually
expressed as limits on the real and imaginary parts of the slepton mass
insertions, i.e. on the properly normalized off-diagonal elements of
$\mathbf{m}_{L}^{2}$, $\mathbf{m}_{E}^{2}$, and $\mathbf{A}_{e}$. As shown
e.g. in Ref.~\cite{MasinaS02}, such limits can be quite tight, with many mass
insertions required to be extremely small. However, this does not tell
anything yet about how natural those small values are. To judge of their
naturality, one should relate them to the flavor-breakings observed in the SM.
Indeed, it is only if a specific flavor-breaking in the slepton sector is
required to be significantly smaller than the known flavor-breakings in the
lepton sector that one can speak of a flavor puzzle.\footnote{Of course, one
remains with the known lepton masses and mixings, whose peculiar structures
are by themselves a flavor puzzle.}

For this purpose, the basis of MFV operators constructed in Section 2 is
optimal. As explained there, the spurion expansions are merely
reparametrizations as long as nothing is imposed on their
coefficients~\cite{ColangeloNS08,NikoPhD}. The essence of MFV, on the other
hand, is to require the size of those coefficients to be at most of
$\mathcal{O}(1)$. Relaxing this constraint, we are back to the full MSSM.
Therefore, it is interesting to translate the current bounds on LFV processes
and EDMs into bounds on the coefficients of the expansions (\ref{MFV}). If a
coefficient is required to be much smaller than one, we can say that it is
fine-tuned. Otherwise, the structures of the soft-breaking terms
$\mathbf{m}_{L}^{2}$, $\mathbf{m}_{E}^{2}$, and $\mathbf{A}_{e}$, though maybe
peculiar, are no less natural (or unnatural) than those of the SM quark and
lepton masses and mixings.%

\begin{table}[t] \centering
$%
\begin{tabular}
[c]{|ll|ll|ll|ll|}\hline
\multicolumn{2}{|c|}{$\mathbf{m}_{L}^{2}$} & \multicolumn{2}{|c|}{$\mathbf{m}%
_{E}^{2}$} & \multicolumn{2}{|c|}{$\operatorname{Re}\mathbf{A}_{e}$} &
\multicolumn{2}{|c|}{$\operatorname{Im}\mathbf{A}_{e}$}\\\hline
$a_{2}\lesssim10^{3}$ & [Masses] & $a_{8}\lesssim10^{3}$ & [Masses] &
$\operatorname{Re}c_{1}\lesssim10^{2}$ & [Stab.] & $\operatorname{Im}%
c_{1}\lesssim2$ & $[d_{e}]$\\
$a_{3}\lesssim10$ & $[\mu\rightarrow e\gamma]$ & $a_{9}\lesssim10^{6}$ &
$[\tau\rightarrow\mu\gamma]$ & $\operatorname{Re}c_{2}\lesssim10^{3}$ &
[Stab.] & $\operatorname{Im}c_{2}\lesssim10^{3}$ & [Stab.]\\
$a_{4}\lesssim10^{4}$ & $[\mu\rightarrow e\gamma]$ & $a_{10}\lesssim10^{9}$ &
$[\tau\rightarrow\mu\gamma]$ & $\operatorname{Re}c_{3}\lesssim10^{3}$ &
$[\mu\rightarrow e\gamma]$ & $\operatorname{Im}c_{3}\lesssim10^{3}$ &
$[\mu\rightarrow e\gamma]$\\
$a_{5}\lesssim10^{3}$ & $[\tau\rightarrow\mu\gamma]$ & $a_{11}\lesssim10^{7}$
& $[\tau\rightarrow\mu\gamma]$ & $\operatorname{Re}c_{4}\lesssim10^{6}$ &
$[\mu\rightarrow e\gamma]$ & $\operatorname{Im}c_{4}\lesssim10^{6}$ &
$[\mu\rightarrow e\gamma]$\\
$a_{6}\lesssim10^{4}$ & $[\mu\rightarrow e\gamma]$ & $a_{12}\lesssim10^{11}$ &
$[\tau\rightarrow\mu\gamma]$ & $\operatorname{Re}c_{5}\lesssim10^{5}$ &
$[\tau\rightarrow\mu\gamma]$ & $\operatorname{Im}c_{5}\lesssim10^{5}$ &
$[\tau\rightarrow\mu\gamma]$\\
$b_{1}\lesssim10^{3}$ & $[\tau\rightarrow\mu\gamma]$ & $b_{4}\lesssim10^{7}$ &
$[\tau\rightarrow\mu\gamma]$ & $\operatorname{Re}c_{6}\lesssim10^{7}$ &
$[\mu\rightarrow e\gamma]$ & $\operatorname{Im}c_{6}\lesssim10^{7}$ &
$[\mu\rightarrow e\gamma]$\\
$b_{2}\lesssim10^{6}$ & $[\tau\rightarrow\mu\gamma]$ & $b_{5}\lesssim10^{10}$
& $[\tau\rightarrow\mu\gamma]$ & $\operatorname{Re}d_{1}\lesssim10^{5}$ &
$[\tau\rightarrow\mu\gamma]$ & $\operatorname{Im}d_{1}\lesssim10^{5}$ &
$[\tau\rightarrow\mu\gamma]$\\
$b_{3}\lesssim10^{8}$ & $[\mu\rightarrow e\gamma]$ & $b_{6}\lesssim10^{13}$ &
$[\tau\rightarrow\mu\gamma]$ & $\operatorname{Re}d_{2}\lesssim10^{8}$ &
$[\tau\rightarrow\mu\gamma]$ & $\operatorname{Im}d_{2}\lesssim10^{8}$ &
$[\tau\rightarrow\mu\gamma]$\\
&  &  &  & $\operatorname{Re}d_{3}\lesssim10^{10}$ & $[\mu\rightarrow
e\gamma]$ & $\operatorname{Im}d_{3}\lesssim10^{10}$ & $[\mu\rightarrow
e\gamma]$\\\hline
\end{tabular}
\ $%
\caption
{Bounds on the slepton soft-breaking terms of the general MSSM, expressed
in terms of the coefficients of the expansions (\ref{MFV}). The $\mathbf
{m}_{L}^2$, $\mathbf{m}_{E}^2$, and $\mathbf{A}_{e}%
$ coefficients are normalized
to $a_1$, $a_7$ and $a_1 a_7$, respectively. To establish those
bounds, the MSSM spectrum (\ref{Gaugino}, \ref{SUSYbreak}) is assumed, as
well as $M_R=10^{12}$ GeV, $\tan\beta=20$, $m_{\nu}= 0$. Finally, the bounds
on $a_{2,8}%
$ come from requiring sleptons to be lighter that $4$ TeV, while those
on $\operatorname{Re}c_{1,2}$ and $\operatorname{Im}c_2$ from the vacuum
stability bounds, approximately enforced as $|c_{1,2}|^2\lesssim
3(a_1+ a_7)$.}
\end{table}%

The bounds on the coefficients are collected in Table~3, along with their
origin. We assume the specific MSSM mass-spectrum of Eqs.~(\ref{Gaugino}) and
(\ref{SUSYbreak}), and take $M_{R}=10^{12}$ GeV, $\tan\beta=20$, and $m_{\nu
}=0$, values for which we expect rather tight bounds given Fig.~1. Further, in
a way similar to how mass-insertions are defined, we normalize the
$\mathbf{m}_{L}^{2}$ coefficients with respect to $a_{1}$, those of
$\mathbf{m}_{E}^{2}$ with respect to $a_{7}$, and those of $\mathbf{A}_{e}$
with respect to the product $a_{1}\times a_{7}$. Only one coefficient is
turned on at a time, in addition to $a_{1}$ and $a_{7}$ which are allowed to
vary between $0.1$ and $8$. This means that contributions requiring
simultaneously two spurion operators are absent. This is sufficient for our
purpose, given the suppression of the spurions when $M_{R}$ and $\tan\beta$
are not too large.

The striking feature of Table~3 is how loose the current experimental bounds
are when translated on the coefficients. As could have been expected from the
strong hierarchies between contributions discussed in the previous section,
only $a_{3}/a_{1}$ and $\operatorname{Im}c_{1}/(a_{1}a_{7})$ are actually
constrained to be of $\mathcal{O}(1-10)$, while all the others can take huge
values. One should also realize that in using the coefficients of the
expansions (\ref{MFV}) as measures of leptonic flavor-violation, one is
performing a rather radical change of basis. Indeed, let us recall that a
fully generic $3\times3$ matrix projected on these MFV basis generates
coefficients spanning several, sometimes even tens of orders of magnitude,
simply because the SM flavor structures used as building blocks are very
special and because the basis operators are nearly aligned. This fact offers
an interesting possibility: if future experimental data requires one or
several coefficients to be much larger than one, it will mean that the MSSM
must contain at least one flavor structure beyond those of the SM, i.e. not
aligned with $\mathbf{Y}_{e}^{\dagger}\mathbf{Y}_{e}$ or $\mathbf{Y}_{\nu
}^{\dagger}\mathbf{Y}_{\nu}$. Such a test could not be performed using the
usual mass insertions.

Another interesting aspect of Table~3 is the dominance of the bounds from the
$\tau\rightarrow\mu\gamma$ mode for a majority of coefficients, especially in
the $\mathbf{m}_{E}^{2}$ sector. On the contrary, $d_{e}$ is at present
competitive only for the flavor-blind phase $\operatorname{Im}c_{1}$ while
$d_{\mu}$ and $d_{\tau}$ are simply absent. Further, for several entries, the
experimental bounds are actually unable to say anything, and we quote in
Table~3 the order of magnitude of the bounds one gets restricting all slepton
masses to be below $4$ TeV, as well as from the stability of the potential
(which we approximately impose as $|c_{1,2}|^{2}\lesssim3(a_{1}+a_{7}%
)$)~\cite{CCBUFB}. This dominance of the bounds on LFV transitions over those
on EDMs can be understood as follows. First, by turning on $\operatorname{Im}%
c_{3}$, one actually turns on a whole set of mass insertions. When both
$\operatorname{Im}c_{3}$ and $a_{3}$ are of $\mathcal{O}(1)$, the $\delta
_{LR}$ mass-insertion contributions to LFV modes are subleading, but since we
turn on only one operator at a time, they now dominate. Second, it is clear
from Fig.~1 that $\operatorname{Im}c_{3}$ generates quite small EDMs, far from
their current experimental bounds. Therefore, what Table~3 actually shows is
that when $\operatorname{Im}c_{3}$ increases by orders of magnitude, one
reaches the bound on $\mu\rightarrow e\gamma$ before that on $d_{e}$.

Overall, this analysis clearly demonstrates that in the lepton sector, the
current experimental sensitivity is still far from the level required to
actually probe non-natural flavor structures, at least in the MFV sense. In
other words, even if the bounds on the mass insertions show peculiar patterns,
they should not be interpreted as posing a fine-tuning problem yet, because
they can all be understood naturally in terms of the patterns observed in the
lepton masses and mixings.

\section{Conclusions}

In the present paper, we have constructed the most general expansions for the
slepton soft-breaking terms $\mathbf{m}_{L}^{2}$, $\mathbf{m}_{E}^{2}$, and
$\mathbf{A}_{e}$ within the MSSM, assuming Minimal Flavor Violation and a
seesaw mechanism of type I. All the possible CP-violating phases were
introduced, classified, and their impact on leptonic EDMs analyzed. Our main
results may be summarized as follows:

\begin{enumerate}
\item The occurrence of complex phases in MFV has been thoroughly examined. We
have proved that in principle, the MFV coefficients can never be assumed to
respect CP when CP-violating phases are present in the Yukawa couplings.
Indeed, these coefficients are always defined up to some phases proportional
to the Jarlskog invariant (\ref{Jarlskog}). On one hand, this result implies
that limiting the CP-violating sources within MFV to those present already in
the SM cannot be based on a symmetry principle, and is thus instead a kind of
fine-tuning. On the other hand, it shows that if one can live with this
fine-tuning or if one has a mechanism able to enforce it, it is a numerically
stable situation thanks to the smallness of the Jarlskog invariant. We have
argued that ultimately, it is by comparing with experiment that one can decide
whether such a fine-tuning is present or not, and thus that for now, MFV
coefficients should be taken complex. This conclusion is fully general: it
applies to both the lepton and quark sectors, and whether MFV is imposed on
the MSSM or used model-independently to parametrize generic New Physics
operators~\cite{DambrosioGIS02,CiriglianoGIW05,CiriglianoG06,HurthIKM08}.

\item We have characterized the CP-violating phases by ordering them into
three classes: flavor-blind, flavor-diagonal and flavor off-diagonal;
according to their hierarchical, decreasing impacts on the EDMs (see Fig.~1).
Interestingly, because the MFV operator basis we constructed is hermitian, all
the CP-violating phases coming from the spurion (i.e., from $\mathbf{Y}_{\nu}%
$) are of the flavor off-diagonal type: they have a negligible impact on the
EDMs because they start to contribute only at the second order in the
mass-insertion approximation. Also, to a good approximation, only one MFV
operator is relevant for each type of phases contributing to $d_{e}$, while
$d_{\mu}$ and $d_{\tau}$ can receive additional contributions, especially at
large $\tan\beta$. Even if the ratios of EDMs scale as the lepton masses only
in the presence of the flavor-blind phase, no scenario can enhance $d_{\mu}$
or $d_{\tau}$ sufficiently to make them more promising than $d_{e}$ for
finding a New Physics signal in the near future. Finally, though we
concentrated exclusively on the slepton sector, our classification of the
CP-violating phases can be immediately applied to the quark sector, for which
the MFV expansions can also be written entirely in terms of hermitian operators.

\item For a realistic range of MSSM and neutrino parameters, such that
$B(\mu\rightarrow e\gamma)$ satisfies its experimental bound, the three types
of phases are allowed by the current bound on $d_{e}$. The next generation of
experiments searching for $d_{e}$ should cover most of the range of values
induced by the flavor-blind phase of $\mathbf{A}_{e}$, but will probably not
probe the flavor-diagonal or off-diagonal phases yet. Still, without a signal,
the flavor blind phase of $\mathbf{A}_{e}$ would pose the same problem as the
other flavor blind phases, i.e. those of $\mu$, $M_{1}$ or $M_{2}$, whose
sizes already have to be tightly constrained to pass the current bound on
$d_{e}$ when sparticles are not extremely heavy. A similar conclusion was
reached recently in Ref.~\cite{AltmannshoferBP08}, including also the
flavor-blind phases of the squark sector, though in a slightly more
constrained scenario than MFV.

\item When the general MSSM soft-breaking terms are projected on the MFV
spurion basis, the coefficients can a priori span several orders of magnitude,
instead of $\mathcal{O}(1)$ values within MFV~\cite{ColangeloNS08}. We have
argued that translating the experimental limits into bounds on these
coefficients permits to unambiguously appreciate the amount of fine-tuning
involved in the flavor sector, something the usual mass-insertions do not
immediately permit. Indeed, it is respectively to the SM flavor-breaking that
those of the MSSM should be compared. We have performed such an analysis in
the slepton sector, and found that current experimental bounds are far from
the MFV level, except for two operators whose coefficients are already
constrained to be of $\mathcal{O}(1)$ by $\mu\rightarrow e\gamma$ and $d_{e}$.
Also, the $\tau\rightarrow\mu\gamma$ mode emerges as the most promising to
improve the various limits in this framework. It would be necessary to extend
this analysis to the quark sector to fully appreciate the level at which MFV
is currently tested, and on which observables to concentrate in order to gain
the most information.
\end{enumerate}

In conclusion, the perspectives for a richer CP-violating phenomenology within
the MFV framework are very promising. In particular, the study of the quark
sector initiated in Ref.~\cite{AltmannshoferBP08} should be extended into a
full-fledged MFV analysis, with potentially interesting results for $K$ and
$B$ physics
phenomenology~\cite{Pheno1,Pheno2,ColangeloNS08,HurthIKM08,CERNreportQuark}.
Also, in the MSSM without R-parity, the rather small value for $M_{R}$ implied
by $B(\mu\rightarrow e\gamma)$ means that the bounds on the proton lifetime
are quite easy to satisfy with the help of MFV
alone~\cite{NikolidakisS07,ProcRPV}. This setting then predicts significant
CP-violating and baryon number violating couplings, whose phenomenological
implications have not yet been fully explored. Finally, leptogenesis within
the MFV framework has led to interesting recent
developments~\cite{CiriglianoIP06,BrancoEtAl06,Ciriglianoetal07}, and the
impacts of the additional CP-violating phases are yet to be studied in that context.

\subsection*{Acknowledgements}

We would like to thank Gilberto Colangelo for his comments and support. C. S.
would also like to thank Tobias Hurth for numerous discussions about the issue
of CP-violating phases in the MFV framework, as well as the University of Bern
for its hospitality. This work is partially supported by the EU contract No.
MRTN-CT-2006-035482, \textit{FLAVIAnet}, by the Schweizerischer Nationalfonds,
and by project C6 of the DFG Research Unit SFT-TR9
\textit{Computergest\"{u}tzte Theoretische Teilchenphysik}. The Center for
Research and Education in Fundamental Physics (University of Bern) is
supported by the \textit{Innovations- und Kooperationsprojekt C13} of the
\textit{Schweizerische Universit\"{a}tskonferenz SUK/CRUS}.

\end{document}